\documentclass{article}%
\usepackage{amsmath,amsthm}%
\usepackage{amsfonts, color}%
\usepackage{amssymb}%
\usepackage{graphicx}
\usepackage{verbatim} 
\usepackage[T1]{fontenc} 
\usepackage[utf8]{inputenc} 
\usepackage{authblk} 
\usepackage[numbers]{natbib}
\usepackage{booktabs}
\usepackage{subcaption}
\usepackage{multirow}

\usepackage[margin=1in]{geometry}

\newcommand{\bdsm}{\boldsymbol}

\begin{document}

\title{Tree-based Regression for Interval-valued Data}
\author[1]{Chih-Ching Yeh}
\author[2]{Yan Sun\thanks{Corresponding author. Email: yan.sun@usu.edu}} 
\author[3]{Adele Cutler}
\affil[1,2,3]{Department of Mathematics \& Statistics\\  
Utah State University\\
3900 Old Main Hill\\
Logan, Utah 84322} 

\date{}
\maketitle

\begin{abstract}
Regression methods for interval-valued data have been increasingly studied in recent years. As most of the existing works focus on linear models, it is important to note that many problems in practice are nonlinear in nature and therefore development of nonlinear regression tools for interval-valued data is crucial. In this paper, we propose a tree-based regression method for interval-valued data, which is well applicable to both linear and nonlinear problems. Unlike linear regression models that usually require additional constraints to ensure positivity of the predicted interval length, the proposed method estimates the regression function in a nonparametric way, so the predicted length is naturally positive without any constraints. A simulation study is conducted that compares our method to popular existing regression models for interval-valued data under both linear and nonlinear settings. Furthermore, a real data example is presented where we apply our method to analyze price range data of the Dow Jones Industrial Average index and its component stocks. 
\end{abstract}


\section{Introduction}\label{intro}
Regression with interval-valued data has been attracting increasing interest among researchers. There are various models built upon set arithmetic, which typically view interval-valued data as realizations of one-dimensional random sets (\cite{Diamond90}, \cite{Korner98}, \cite{Gil02}, \cite{Gil07}, \cite{G-R07}, \cite{Blanco11}, \cite{Cattaneo12}). Separately, there are also plentiful models developed under symbolic data analysis (SDA) that treat an interval as a bivariate vector (\cite{Carvalho04}, \cite{Billard07}, \cite{Neto08}, \cite{Neto10}). A detailed review of these models will be given in Section \ref{sec:regmethod}. Generally speaking, models from both domains have their distinct advantages and disadvantages, but they all share the common drawback of being too restrictive. In other words, all of these models are rigid to some degree due to the constraints they need to impose on their parameters. This is rooted in the fundamental fact that the space of intervals $\mathcal{K}_{\mathcal{C}}(\mathbb{R})$ as a metric space is not linear (see \ref{append:preliminary}). In addition, many data in practice are nonlinear, for which linear models are insufficient. Motivated by all of these, we propose to study regression of interval-valued data by nonparametric approaches. Without any distribution assumption, nonparametric methods usually do not require additional constraints and therefore are more flexible. Furthermore, they can take care of both linearity and nonlinearity. Particularly in this paper, we propose a tree-based regression method based on random forests.  

Random forests (\cite{Breiman01}) are ensembles of classification or regression trees created using bootstrap samples of the training data and random feature selection in tree induction. The method is among the most popular algorithms and widely used in data science and machine learning in appreciation of its high prediction performance. Unlike other ensemble methods, random forests involve additional randomness by selecting a random subset of the predictor variables at each splitting node, which makes it robust to overfitting and high-dimensionality. For classification, the prediction is made by the majority vote from all the trees. For regression, the prediction is instead determined by the average of the tree predictions.  For interval-valued data, comparing to other regression methods, random forests automatically overcome the difficulty of mathematical coherence, i.e., the predicted interval range/radius must be nonnegative, due to their nonparametric nature. Thus, it is expected to perform better, especially when data do not fulfill the restrictions of other regression methods. Furthermore, the fact that random forests can deal with both linearity and nonlinearity makes them more flexible methods.
	 
The rest of the paper is organized as follows. Section \ref{sec:regmethod} provides a review of major (linear and nonlinear) regression methods for interval-valued data in the literature, where in particular an extension of the kernel method for nonlinear interval regression is discussed. Section \ref{sec:RF} describes random forests as classical methods of classification and regression in machine learning, and proposes their interval-valued adaptation. Results of a systematic simulation study that compares random forest regression to typical existing methods are reported in Section \ref{sec:simulation}, and real data applications are presented in Section \ref{sec:application}. Section \ref{sec:conclude} concludes with remarks for future research. Preliminaries of random sets theory are deferred to the Appendix. 
 
Throughout the paper, we denote by $[x]\in\mathcal{K}_{\mathcal{C}}(\mathbb{R})$ a bounded closed interval, whose lower and upper bounds are denoted by $x^L$ and $x^U$, respectively. Alternatively, $[x]$ can also be represented by its center and radius, denoted by $x^C$ and $x^R$, respectively.  A random interval which takes values in $\mathcal{K}_{\mathcal{C}}(\mathbb{R})$ is denoted by $[X]$. Bolded letters denote vectors. For example, $[\bdsm{x}]=\left[[x_1], \cdots, [x_p]\right]^T$ denotes a $p$-dimensional hyper interval, and its random version is denoted by $[\bdsm{X}]$. There are various metrics defined for the space $\mathcal{K}_{\mathcal{C}}(\mathbb{R})$ (see \ref{append:preliminary}), which usually can be used for the same model with the choice up to the user. We adopt a unifying notation $d(\cdot, \cdot)$ for all of the metrics. 

\section{Review of Regression Methods}\label{sec:regmethod}
\subsection{Linear regression}
Linear regression for interval-valued data has been extensively studied in the past decades. Existing models have been developed mainly in the two domains of random sets and symbolic data analysis (SDA). In the framework of random sets, an interval is viewed as a compact convex set in the one-dimensional real space $\mathbb{R}$, and the linear relationships between intervals are modeled according to set arithmetic. Separately, the aim of SDA is to extend classical data analysis techniques to nontraditional data formats, such as lists, intervals, histograms and distributions. In this framework, linear regression is often developed by fitting separate point-valued models to the center and radius (or the lower and upper bounds), essentially treating an interval as a bivariate vector. In the following, we briefly review major models from both domains. 

The foundation of linear regression in the random sets framework is the linear structure in the space $\mathcal{K}_{\mathcal{C}}(\mathbb{R}^d)$ defined by the Minkowski addition and scalar multiplication, i.e.
\begin{eqnarray*}
  &&A+B=\left\{a+b: a\in A, b\in B\right\},\\
  &&\lambda A=\left\{\lambda a: a\in A\right\},\\ 
  &&A, B\in\mathcal{K}_\mathcal{C},\ \ \lambda\in\mathbb{R}.
\end{eqnarray*}
For $[x], [y]\in\mathcal{K}_{\mathcal{C}}(\mathbb{R})$, this means  
\begin{eqnarray}
  &&[x]+[y]=[x^L+y^L, x^U+y^U],\label{def:int_add}\\
  &&\lambda [x]=
  \begin{cases}
  [\lambda x^L, \lambda x^U], &\lambda\geq 0,\\
  [\lambda x^U, \lambda x^L], &\lambda<0.
  \end{cases}\label{def:int_multi}
\end{eqnarray}
Although different researchers may have different notation and degrees of restrictions (\cite{Gil01}, \cite{Gil02}, \cite{Gil07}, \cite{G-R07}, \cite{Sinova12}), linear regression models in $\mathcal{K}_{\mathcal{C}}(\mathbb{R})$ generally have the form
\begin{equation}\label{def:mod_set}
  [Y]=a[X]+[\mathcal{E}],
\end{equation}
where $a\in\mathbb{R}$ and $[\mathcal{E}]$ is an interval-valued random error with a fixed expectation $E\left([\mathcal{E}]\right)=[b]\in\mathcal{K}_{\mathcal{C}}(\mathbb{R})$.
Here the addition and multiplication are in the sense of (\ref{def:int_add}) and (\ref{def:int_multi}), respectively. It follows that 
\begin{equation}\label{pred:mod_set}
  \widehat{[Y]}=E\left([Y]\mid[X]\right)=a[X]+[b],
\end{equation}
Equivalently, the equation (\ref{pred:mod_set}) can be written in terms of the center and radius as
\begin{eqnarray}
  \hat{Y}^C&=&aX^C+b^C,\label{pred:mod_set_c}\\
  \hat{Y}^R&=&|a|X^R+b^R. \label{pred:mod_set_r}
\end{eqnarray}
This leads to the following equation that shows linearity in $\mathcal{K}_{\mathcal{C}}$:
\begin{equation}\label{eqn:linearity}
  d\left(\widehat{[Y_1]}, \widehat{[Y_2]}\right)=|a|d\left([X_1], [X_2]\right).
\end{equation}
That is, a constant change in the metric of $[X]$ results in a constant change in the metric of $[Y]$. This reveals the essence of model (\ref{def:mod_set}) as an analogous result of the fundamental property of classical linear regression. However, such a property has to be achieved at the price of reduced model flexibility from practical point of view. Notice from (\ref{pred:mod_set_c})-(\ref{pred:mod_set_r}) that the slopes for the center and radius equations must have the same absolute value. This reduction of flexibility is the price we pay to achieve linearity in $\mathcal{K}_{\mathcal{C}}$. 

The multivariate extension of model (\ref{def:mod_set}) is given as
\begin{eqnarray}\label{def:mod_set_multi}
  [Y]=\sum_{i=1}^{p}a_i[X_i]+[{\mathcal{E}}],
\end{eqnarray}
or equivalently, in the center-radius form
\begin{eqnarray}
  Y^C&=&\sum_{i=1}^{p}a_iX_i^C+\mathcal{E}^C,\label{mod_set_c}\\
  Y^R&=&\sum_{i=1}^{p}|a_i|X_i^R+\mathcal{E}^R, \label{mod_set_r}
\end{eqnarray}
where $a_i\in\mathbb{R}$, $i=1,\cdots, p$, and $\mathcal{E}^C, \mathcal{E}^R$ are random variables with $E(\mathcal{E}^C)=b^C\in\mathbb{R}$ and $E(\mathcal{E}^R)=b^R>0$. Estimation of the model parameters is generally performed by the least squares method that minimizes the mean squared errors with respect to the metric in $\mathcal{K}_{\mathcal{C}}(\mathbb{R})$. Assuming observing a sample of size $n$, the least squares estimates are given by 
\begin{equation}
  \left\{\hat{a}_1, \cdots, \hat{a}_p;  \widehat{[b]}\right\}
  =\arg\min \left\{ \frac{1}{n}\sum_{j=1}^{n}d^2\left([y_j], \sum_{i=1}^{p}a_i[x_{ij}]+[b]\right)\right\},
\end{equation}
subject to the constraint $b^R>0$. 

In the separate framework of SDA, interval linear regression has been studied mainly by treating the intervals as bivariate vectors. Models belonging to this category typically specify separate linear equations for the center and radius (or upper and lower bound), respectively. For example, the MinMax method by \cite{Billard02} has the model equations 
\begin{eqnarray*}
  Y^L&=&\beta_0^L+\sum_{i=1}^{p}\beta_i^LX_i^L+\epsilon_1,\ \ \beta_i^L\in\mathbb{R},\ i=0, \cdots, p,\\
  Y^U&=&\beta_0^U+\sum_{i=1}^{p}\beta_i^UX_i^U+\epsilon_2,\ \ \beta_i^U\in\mathbb{R},\ i=0, \cdots, p. 
\end{eqnarray*}
Alternatively, \cite{Neto08} proposed the Center and Range Method (CRM) as
\begin{eqnarray*}
  Y^C&=&\beta_0^C+\sum_{i=1}^{p}\beta_i^CX_i^C+\epsilon_1,\ \ \beta_i^C\in\mathbb{R},\ i=0, \cdots, p,\\
  Y^R&=&\beta_0^R+\sum_{i=1}^{p}\beta_i^RX_i^R+\epsilon_2,\ \ \beta_i^R\in\mathbb{R}.\ i=0, \cdots, p.
\end{eqnarray*}
Here, both $\epsilon_1$ and $\epsilon_2$ are zero-mean random errors. There is no interval structure between them. Obviously, without any constraint, these two methods suffer from the problem of mathematical coherence, i.e., the predicted upper bound may be smaller than the lower bound, or the predicted radius may be negative. This leads to the proposal of the Constrained Center and Range Method (CCRM) (\cite{Neto10}), which takes the same form of CRM, with the additional constraints $\beta_0^R, \beta_1^R\geq 0$ to ensure the nonnegativity of $\hat{Y}^R$. There have been further developments based on this type of modeling. For examples, \cite{Domingues10} considered a parametric inferential method in replacement of the least squares to deal with the issue of outliers, and \cite{Neto11} proposed a generalized linear model to allow for additional model flexibility. Although the bivariate representation of an interval could result in loss of geometric information and less interpretability (e.g., the linearity property (\ref{eqn:linearity}) does not hold anymore), this type of models generally has improved flexibility, and therefore are preferred in some practical situations.  

There is an intrinsic difficulty for linear regression with interval-valued data, because $\mathcal{K}_{\mathcal{C}}(\mathbb{R})$ is not a linear space (see, e.g., \cite{Sun16} for a detailed discussion). More specifically, although the Minkowski addition and scalar multiplication define a linear structure in $\mathcal{K}_{\mathcal{C}}(\mathbb{R})$, there is no inverse element of addition.  This results in the biggest problem for most models to impose non-negative constraints. Technically, such constraints usually mean that underestimation is more heavily penalized than overestimation, leading to biased estimators. They also introduce significant computational complication, making it difficult to draw inferences. 

\subsection{Kernel method for nonlinear regression}
Although linear regression is often preferred for its simplicity, there are many situations in practice where nonlinearity does exist and linear methods alone are not sufficient to tackle those problems. Therefore, developing nonlinear regression methods for interval-valued data is important. However, compared to linear regression, there are very few papers in the literature where nonlinear regression is rigorously studied. Among those, we are particularly interested in the kernel approach (\cite{Jeon15}). It is essentially an extension of the multivariate kernel density estimator to interval-valued data. The main purpose of the method is to estimate the multivariate density function based on interval-valued data. Nevertheless, it can also be used to perform nonlinear regression in an indirect way. After a brief review of the kernel approach proposed in \cite{Jeon15}, we will discuss an immediate extension of this method to directly address nonlinear regression with interval-valued data. 

The classical multivariate kernel density estimator has the form
\begin{equation*}
  \hat{f}(\bdsm{x})=\frac{1}{n}\sum_{j=1}^{n}K_h\left(\bdsm{x}-\bdsm{x}_j\right)/h,
\end{equation*}
where $K_h(\cdot)$ is a multivariate kernel function, $\bdsm{x}=[x_1, \cdots, x_p]^T$ denotes a $p$-dimensional vector, and $h$ is the bandwidth parameter. There is a range of popularly used kernel functions, including uniform, triangular, Epanechnikov, and Gaussian. The bandwidth parameter $h$ controls the smoothness of the estimated density function: smaller values of $h$ can reflect more local structures resulting in a wiggly function, while larger values of $h$ make the function smoother but with possible loss of local details. In \cite{Jeon15}, the interval-valued observation is viewed as a histogram, to which a kernel distribution is fitted. Precisely, their proposed density estimator is defined as 
\begin{equation}\label{eqn:kernel_pdf}
  \hat{f}(\bdsm{x})=\frac{1}{n}\sum_{j=1}^{n}\phi\left(\bdsm{x}|\hat{\bdsm{\mu}}_j, \hat{\bdsm{\Sigma}}_j\right),\ \bdsm{x}\in\mathbb{R}^p,
\end{equation}
where $\phi(\cdot|\bdsm{\mu},\bdsm{\Sigma})$ is the multivariate Gaussian kernel (i.e., multivariate normal density). For the $j^{th}$ interval observation, the correponding mean 
$\hat{\bdsm{\mu}}_j$ and covariance $\hat{\bdsm{\Sigma}}_j$ are estimated from the data with a unique set of weights. In the regression setting with outcome variable $Y$ and predictor variables $X_i$, $i=1,\cdots,p$ (for which interval-valued data is observed), one computes the conditional density of $Y$ given $\bdsm{X}=[X_1,\cdots, X_p]^T$ as
\begin{equation}\label{eqn:kernel_cpdf}
  \hat{f}_{Y|\bdsm{X}}(y|\bdsm{x})=\frac{\hat{f}_{Y, \bdsm{X}}(y, \bdsm{x})}{\hat{f}_{\bdsm{X}}(\bdsm{x})},\ \bdsm{x}\in\mathbb{R}^p,\ y\in\mathbb{R}.
\end{equation}
The drawback of this method is that, although it makes good use of the interval-valued data, it does not develop an interval-valued model. Namely, the densities (\ref{eqn:kernel_pdf}) and (\ref{eqn:kernel_cpdf}) estimated from the interval-valued data are still in the point-valued context. Consequently, prediction for the interval-valued output is somewhat less straightforward. For example, \cite{Jeon15} proposed a rather complicated prediction method, of which the key is to calculate the conditional cumulative distribution function 
\begin{equation*}
  P(Y\leq y|\bdsm{x}^L\leq\bdsm{X}\leq\bdsm{x}^U)
\end{equation*}
based on the estimated conditional density $\hat{f}$. Then, the center is predicted as the conditional expectation, and the lower and upper bounds are predicted by the $\hat{q}_{L}^{th}$ and $\hat{q}_{U}^{th}$ quantiles of the conditional distribution, respectively. The values of $\hat{q}_L$ and $\hat{q}_U$ are estimated by the relative positions of $\bdsm{x}^L$ and $\bdsm{x}^U$ in the sample. 

In fact, in a regression context, it is more natural to directly estimate the conditional mean as a function of the input variables, without the need of the conditional density. For example, we may consider using the kernel approach to estimate the following two regression function for the center and radius, respectively, 
\begin{eqnarray}
  m^C\left(x_i^C, x_i^R\right)&=&E\left(Y^C|X_i^C=x_i^C, X_i^R=x_i^R\right),\label{mod:kernel_c}\\
  m^R\left(x_i^C, x_i^R\right)&=&E\left(Y^R|X_i^C=x_i^C, X_i^R=x_i^R\right).\label{mod:kernel_r}
\end{eqnarray} 
It is in a similar spirit to CCRM, except that we add the radii and centers of the predictor intervals in both the center and radius equations, to make the model more flexible. The kernel estimators for the regression functions at $[\bdsm{X}]=[\bdsm{x}^{*}]$ are
\begin{eqnarray}
  \hat{m}^C\left([\bdsm{x}^*]\right)&=&\frac{\sum_{j=1}^{n}K\left(\frac{d(\bdsm{x}^*, \bdsm{x}_j)}{h}\right)Y_j^C}{\sum_{j=1}^{n}K\left(\frac{d(\bdsm{x}^*, \bdsm{x}_j)}{h}\right)},\label{kernel_c}\\
  \hat{m}^R\left([\bdsm{x}^*]\right)&=&\frac{\sum_{j=1}^{n}K\left(\frac{d(\bdsm{x}^*, \bdsm{x}_j)}{h}\right)Y_j^R}{\sum_{j=1}^{n}K\left(\frac{d(\bdsm{x}^*, \bdsm{x}_j)}{h}\right)},\label{kernel_r}
\end{eqnarray}
where $d(\bdsm{x}^*, \bdsm{x}_j)$ is the generalized distance between two $p$-dimensional hyper intervals defined by 
\begin{eqnarray*}
  d^2(\bdsm{x}^*, \bdsm{x}_j)=\sum_{i=1}^{p}\left\{\left[(x_i^*)^C-x_{i,j}^C\right]^2+\left[(x_i^*)^R-x_{i,j}^R\right]^2\right\}.
\end{eqnarray*}
As for the kernel density estimator, here $K(\cdot)$ is a kernel function and $h$ is the bandwidth parameter. As we can see from (\ref{kernel_c})-(\ref{kernel_r}), the kernel regression function is essentially a weighted average, so the predicted radius $\hat{m}^R(\cdot)$ is automatically positive without any constraints.


\section{Random Forest Regression for Interval-valued Data}\label{sec:RF}
As we discussed previously, linear regression for interval-valued data suffers from the intrinsic problem that $\mathcal{K}_{\mathcal{C}}(\mathbb{R})$ is not a linear space. Consequently, linear models typically have to impose non-negativity constraints. For example, model (\ref{mod_set_c})-(\ref{mod_set_r}) needs to have a positive expectation for $\mathcal{E}^R$. In CCRM, all of the coefficients in the radius equation $\beta_i^R, i=0,1,\cdots,p$, are constrained to be nonnegative. Therefore, much of classical linear regression theory, such as least squares, does not apply. Instead, a constrained optimization algorithm is needed, making inference difficult. In the preceding section, we discussed the possibility of kernel regression for interval-valued data. Given its nonparametric nature, it has the advantage of solving the problem of mathematical coherence automatically. Furthermore, it can handle nonlinearity. However, the kernel method suffers from the curse of dimensionality, i.e., when the number of predictors gets large, the method tends to be slow and inaccurate (e.g., \cite{Linton95}). In addition, the performance of the kernel method depends heavily on the choice of the kernel function and the bandwidth parameter, so tuning is a big issue. In order to address these drawbacks of the kernel method, while still keeping its advantages, we propose a random forests regression for interval-valued data. 

The random forest algorithm is an ensemble method developed by Leo Breiman (\cite{Breiman01}). A random forest is an ensemble of a collection of trees. Before the introduction of random forests, ensembles had already attracted lots of attention, as they were observed to achieve much more accurate predictions than individual trees. Many authors contributed various techniques for constructing ensembles. Among these, popular examples include Bagging (\cite{Breiman96}), Adaboost (\cite{Freund96}), and Randomization (\cite{Dietterich00}). In Bagging, each tree is constructed from a bootstrap sample drawn with replacement from the training data. The original version of Adaboost resamples observations with weights that are successively adjusted to give higher weight to ``difficult'' observations. Randomization generates an ensemble by randomizing the interval decisions made by the base algorithm. After the ensembles are constructed, the majority vote of the individual classifiers is taken for classification and the average is taken for regression. It is well understood that averaging results in variance reduction, and reducing the correlation between individual classifiers further enhances the variance gains (\cite{Breiman01}). Motivated by this principle, random forests injects another layer of randomness by changing the structure of each tree. Instead of optimizing the response by evaluating all the predictors, as is done with single-tree methods or bagging, a subset of the predictors, drawn at random independently for each node in each tree, is employed. This strategy turns out to perform very well and is robust against overfitting.

One of the main advantages random forests have over other estimation methods is that they are fully non-parametric, including the effects of the predictors and response variables. So they well handle the issues of nonlinearity and mathematical coherence. Precisely, because the radius is predicted essentially by an average of the terminal nodes, which contain all positive elements, i.e., radii of the observed intervals, the prediction is automatically positive without any constraints. Situations where there are as many or more predictor variables than observations may also be problematic for traditional methods, but random forests tackle the issue of dimensionality automatically, largely due the use of decision trees as the base learners (a building block for an ensemble process). Like all tree-based methods, random forests automatically fit interactions without the interacting predictors needing to be specified a priori. Finally and importantly, the random forests algorithm only has three tuning parameters: the size of the subset at each node, the number of trees in the forest, and the depth of the trees, which makes it conveniently applicable in practice. In fact, the number of trees can be chosen as large as desired without risk and for classification, the trees can be grown to the deepest possible depth. The results are not particularly sensitive even to the size of the subset at each node (\cite{Breiman01}), so tuning is relatively straightforward.

For random forests regression with interval-valued data, we propose the same regression equations (\ref{mod:kernel_c})-(\ref{mod:kernel_r}) as in the kernel method. For completeness, we describe the basic random forests regression algorithm in the following. Readers are referred to \cite{Liaw-Wiener02}) for more details. 
\begin{enumerate}
\item For each of the $N$ desired regression trees, draw a bootstrap sample from the original data.
\item For each bootstrap sample, grow a tree as described in steps 3 and 4. 
\item At each node, select \textit{p/3} predictor variables and find all possible splits on these predictors. The ``best split'' is determined by minimizing the residual sum of squares over all splits.
\item Continue splitting until each node has no less than 5 observations. Once a node reaches this criterion, the node is said to be a terminal node.
\item To make a prediction for a new observation, drop the observation down each of the $N$ trees and average the predictions. 
\item The prediction accuracy is further enhanced by using the data not in the bootstrap as the test data. The estimate of the error rate is acquired by aggregating this out-of-bag (OOB) predictions, and is called the OOB estimate of error rate.  
\end{enumerate}


\section{Simulation}\label{sec:simulation}
In this section, we examine the empirical performance of random forest regression and compare it to CCRM and kernel estimator. For these purposes, we simulate four settings of linear and three settings of non-linear data. For each setting, three datasets with n = 500,  1000 and 2000 rectangles are generated. Of each dataset, 10\% is used as the training set and the remaining 90\% is used as the test set. For example, the dataset of size n = 500 is split into a training set and test set with n = 50 and 450, respectively. The models are fit to the training sets, and then the prediction errors are computed on the test sets. The predictive accuracy is measured by three popularly used criteria: coefficient of determination ($R^{2}$), 
mean squared error (MSE) and mean absolute error (MAE). The process is repeated for each setting/sample size combination 100 times independently, and the average results are reported in tables \ref{tab:LRc}-\ref{tab:NONr}. All of our analysis is performed in R. The random forests algorithm is implemented using the \textit{randomForest} package, and the kernel regression and CCRM are implemented using the packages \textit{np} and \textit{iRegression}, respectively. 
Details of our simulation settings are described below.
\begin{itemize}
		\item Setting 1: Linear model (\ref{pred:mod_set_c})-(\ref{pred:mod_set_r}) with $r>0$. 
	The center of the predictor interval $X^{C}$ is generated from $N(5, 2^{2})$, and the radius $X^{R}$ is generated from $U(0.5, 1.5)$, both independently. The center and radius of the response interval are determined by the equations
	\begin{eqnarray*}
	  &&Y^{C}_{i} = 2X^{C}_{i} + 5 + \epsilon_i^C,\\
	  &&Y^R_i = 2X^{R}_{i} + \epsilon_i^R,
	\end{eqnarray*}
	where $\epsilon_i^C \sim N(0, 2^{2})$ and $\epsilon_i^R \sim N(0.5, 0.3^{2})$, $i=1,\cdots, n$. A simulated data set of $n=200$ from setting 1 is shown in figure \ref{setting1_big}.
\begin{figure} 
	\centering
	\includegraphics[width=350pt]{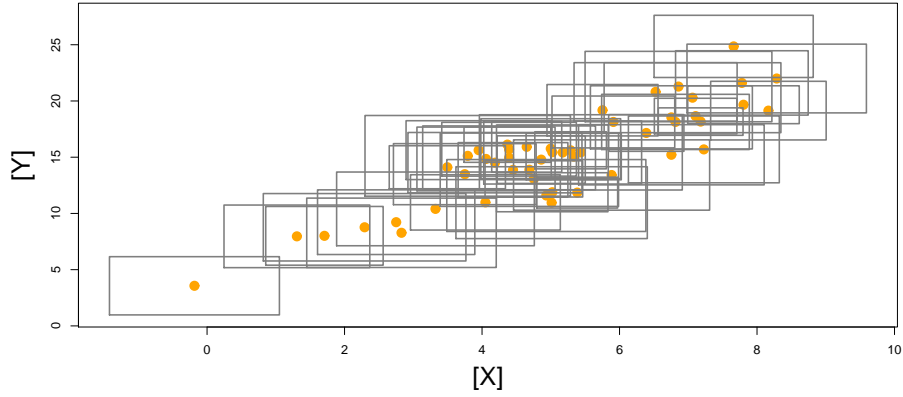}
	\caption{Plot of a simulated data set from setting 1. The gray rectangles denote the interval-valued data, and the yellow dots are the corresponding centers (same for figures \ref{setting2_big}-\ref{setting6_big} in the following).}
	\label{setting1_big}
\end{figure} 
	
		\item Setting 2: Linear model (\ref{pred:mod_set_c})-(\ref{pred:mod_set_r}) with $r<0$. 
	The center of the predictor interval $X^{C}$ is generated from $U (0, 20)$, and the radius $X^{R}$ is generated from $U (10, 11)$, both independently. The center and radius of the response interval follow the same equations as in Setting 1, but $\epsilon_i^C \sim N(0, 5^{2})$ and $\epsilon_i^R \sim N(-15, 0.5^{2})$, $i=1,\cdots, n$. A simulated data set of $n=200$ from setting 2 is shown in figure \ref{setting2_big}.
\begin{figure}
	\centering
	\includegraphics[width=350pt]{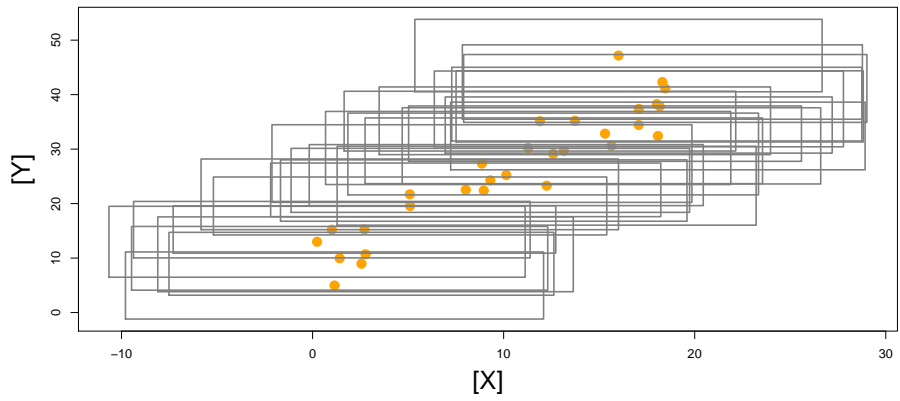}
	\caption{Plot of a simulated data set from setting 2.}
	\label{setting2_big}
\end{figure} 
	
		\item Setting 3: Linear center and radius relationships.  
	The center and radius of the predictor interval are independently generated from $N(5, 5^{2})$ and $U(10, 15)$, respectively. The response intervals are generated according to the equations
	\begin{eqnarray*}
		&&Y^C_i = 10X^{C}_{i} + 20X^{R}_{i} + \eta + \epsilon_i^C,\\
		&&Y^R_i = 2X^{R}_{i} + \theta + \epsilon_i^R,
	\end{eqnarray*}
	where $\eta \sim U(0, 4)$, $\epsilon_i^C \sim N(-5, \sigma^{2})$, $\sigma \sim U(3, 4)$ and $\frac{n}{50}\theta \sim U(0, 2)$, $\epsilon_i^R \sim N(-15, 1^{2})$, $i=1,\cdots, n$. A simulated data set of $n=200$ from setting 3 is shown in figure \ref{setting3_big}.
\begin{figure}
	\centering
	\includegraphics[width=350pt]{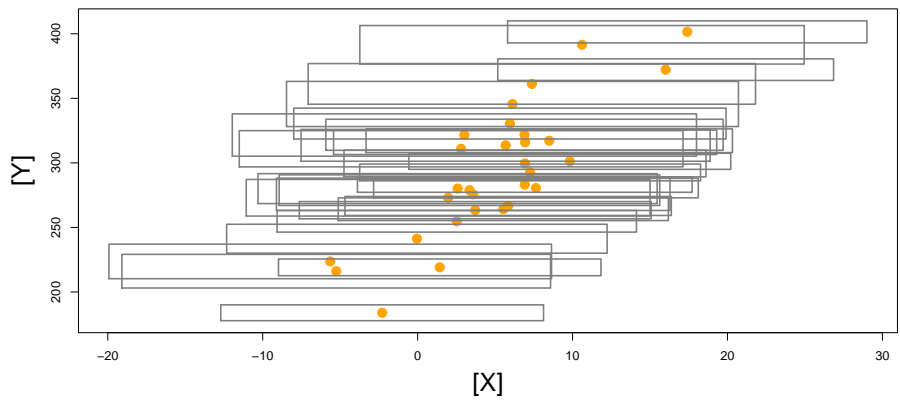}
	\caption{Plot of a simulated data set from setting 3.}
	\label{setting3_big}
\end{figure} 
	
		\item Setting 4: Close-to-linear center and radius relationships.
	The center and radius of the predictor interval are independently generated from $N(5, 0.9^{2})$ and $N(5, 10^{2})$, respectively. The response intervals are determined by
	\begin{eqnarray*}
		&&Y^C_i = 0.22e^{X^{C}_{i}} + \epsilon_i^C,\\
		&&Y^R_i = \Phi(X^R_i; 2, 2)+ \epsilon_i^R,
	\end{eqnarray*}
	where $\Phi(\cdot; \mu, \sigma)$ is the CDF (cumulative distribution function) of $N(\mu, \sigma^2)$, and the random errors are generated by $\epsilon_i^C \sim N(0, \sigma^{2}_C)$, $\sigma^2_C \sim U(15, 20)$, $\epsilon_i^R \sim N(1, \sigma^{2}_R)$, $\sigma^2_R \sim U(0, 1)$, $i=1,\cdots, n$. A simulated data set of $n=200$ from setting 4 is shown in figure \ref{setting4_big}.
\begin{figure}
	\centering
	\includegraphics[width=350pt]{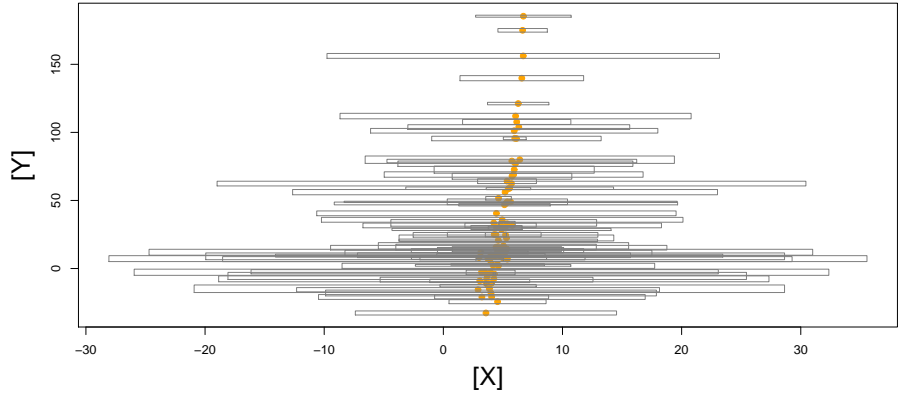}
	\caption{Plot of a simulated data set from setting 4.}
	\label{setting4_big}
\end{figure} 
	
		\item Setting 5: Nonlinear center relationship and linear radius relationship.
	The center and radius of the predictor interval are independently generated from $N(5, 2^{2})$ and $U(0.5, 1.5)$, respectively. The center and radius of the response interval follow the equations
	\begin{eqnarray*}
		&&Y^C_i = 6 + 4\sin{\left(0.25\pi X^{C}_{i}\right)} + \epsilon_i^C,\\
		&&Y^R_i = X^{R}_{i} + 0.5 +  \epsilon_i^R,
	\end{eqnarray*}
	where $\epsilon_i^C \sim N(0, 0.5^{2})$ and $\epsilon_i^R \sim N(0, 0.2^{2})$, $i=1,\cdots, n$. A simulated data set of $n=200$ from setting 5 is shown in figure \ref{setting5_big}.
	\begin{figure} 
		\centering
		\includegraphics[width=350pt]{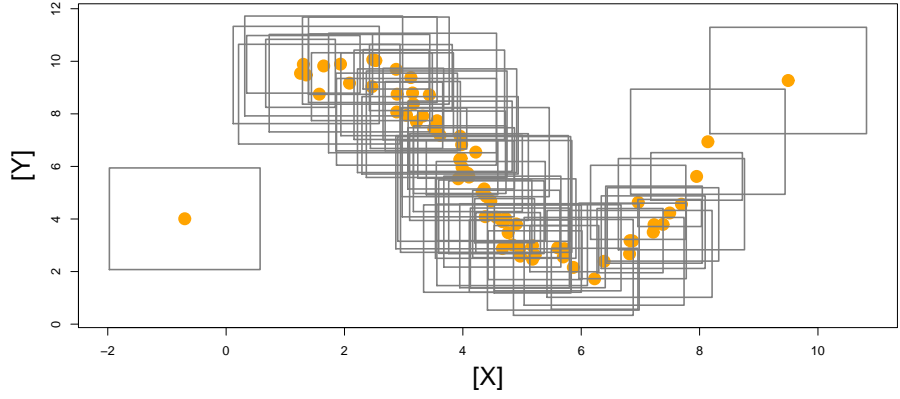}
		\caption{Plot of a simulated data set from setting 5.}
		\label{setting5_big}
	\end{figure} 
	\leavevmode
	\newline
	
		\item Setting 6: Nonlinear center and radius relationships.
	The center and radius of the predictor interval are independently generated from $N(5, 2^{2})$ and $U(0.25, 0.5)$, respectively. The response intervals are generated by 
	\begin{eqnarray*}
		&&Y^C_i = 6 + 2X^{R}_{i} + \sin{\left(0.253\pi X^{C}_{i}\right)} + \epsilon_i^C,\\
		&&Y^R_i = |-0.3X^{C}_{i}X^{R}_{i} + 0.5|+ \epsilon_i^R,
	\end{eqnarray*}
	where $\epsilon_i^C \sim N(0, 0.5^{2})$ and $\epsilon_i^R \sim N(0, 0.1^{2})$, $i=1,\cdots, n$. A simulated data set of $n=200$ from setting 6 is shown in figure \ref{setting6_big}.
\begin{figure}
	\centering
	\includegraphics[width=350pt]{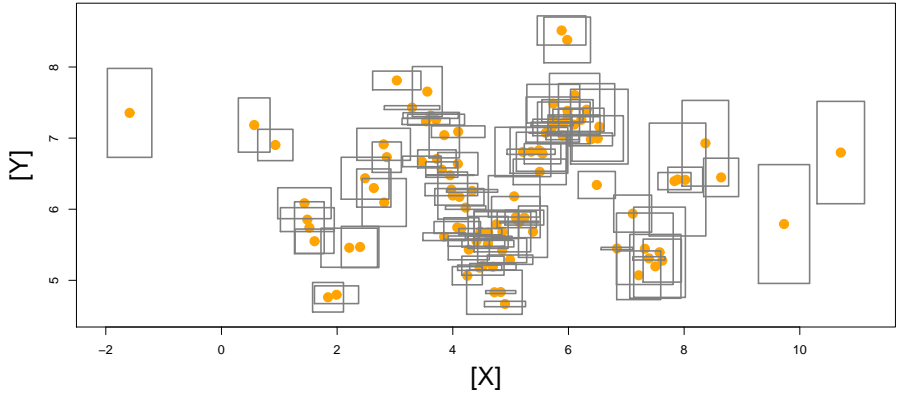}
	\caption{Plot of a simulated data set from setting 6.}
	\label{setting6_big}
\end{figure} 
	
	\item Setting 7: Multivariate nonlinear center and radius relationships with noise variables. 
	There are a total of 5 predictor intervals. Their centers are generated by $X^{C}_{1} \sim N(5, 3^{2})$, $X^{C}_{2} \sim \beta(0.5, 0.5)$, $X^{C}_{3} \sim N(10, 3.5^{2})$, $X^{C}_{4} \sim U(0.5, 1.5)$, $X^{C}_{5} \sim N(8, 3.5^{2})$, 
all independently. Define
\begin{eqnarray*}
  V_{1} &=& u_{1} + e^{-0.5\gamma(3, 2) + \tau_{1}},\\
  V_{2} &=& u_{2} + e^{-0.5\beta(1, 3) + \tau_{2}},
\end{eqnarray*}	
where $\tau_{1}, \tau_{2} \sim N(0, 0.2^{2})$ and $u_{1}, u_{2} \sim U(0, 0.5)$, all independently. The radii of the predictor intervals are generated by $X^{R}_{1} = \frac{2V_{1}}{1 + V_{1}}$, $X^{R}_{2} = \frac{3V_{2}}{1 + V_{2}}$, $X^{R}_{3} \sim N(10, 3^{2})$, $X^{R}_{4} \sim U(2.5, 3.5)$, $X^{R}_{5} \sim \beta(2, 5)$, 
all independently. The center and radius of the response interval are determined by the equations
	\begin{eqnarray*}
		&&Y^C_i = {(X^C_{1i} + (X^C_{1i})^2)}{(X^C_{2i} + (X^C_{2i})^2)} - {(X^C_{3i} + (X^C_{3i})^2)}{(X^C_{4i} + (X^C_{4i})^2)} - X^C_{5i} + \epsilon_i^C,\\
		&&Y^R_i = \frac{(X^{R}_{2i})^{2}}{5} + 0.1X_{3i}^R - 5(X^R_{1i}X^R_{4i} + X^R_{5i}) + 4 + \epsilon_i^R,
	\end{eqnarray*}
where $\epsilon_i^C \sim N(0, 1^{2})$ and $\epsilon_i^R \sim N(-3, 0.15^{2})$, $i=1,\cdots, n$.
\end{itemize}

\begin{figure}
	\centering
	\includegraphics[width=170pt, height=110pt]{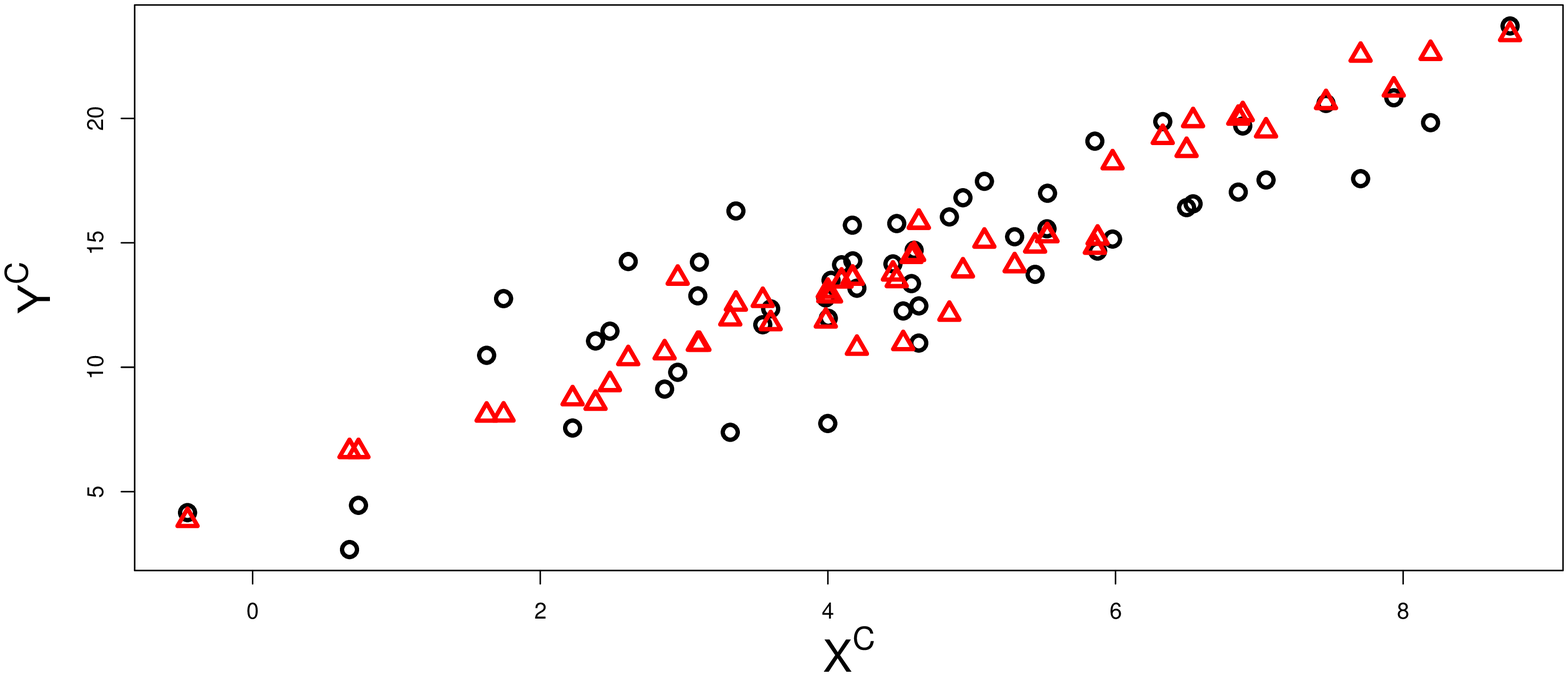} 
	\includegraphics[width=170pt, height=110pt]{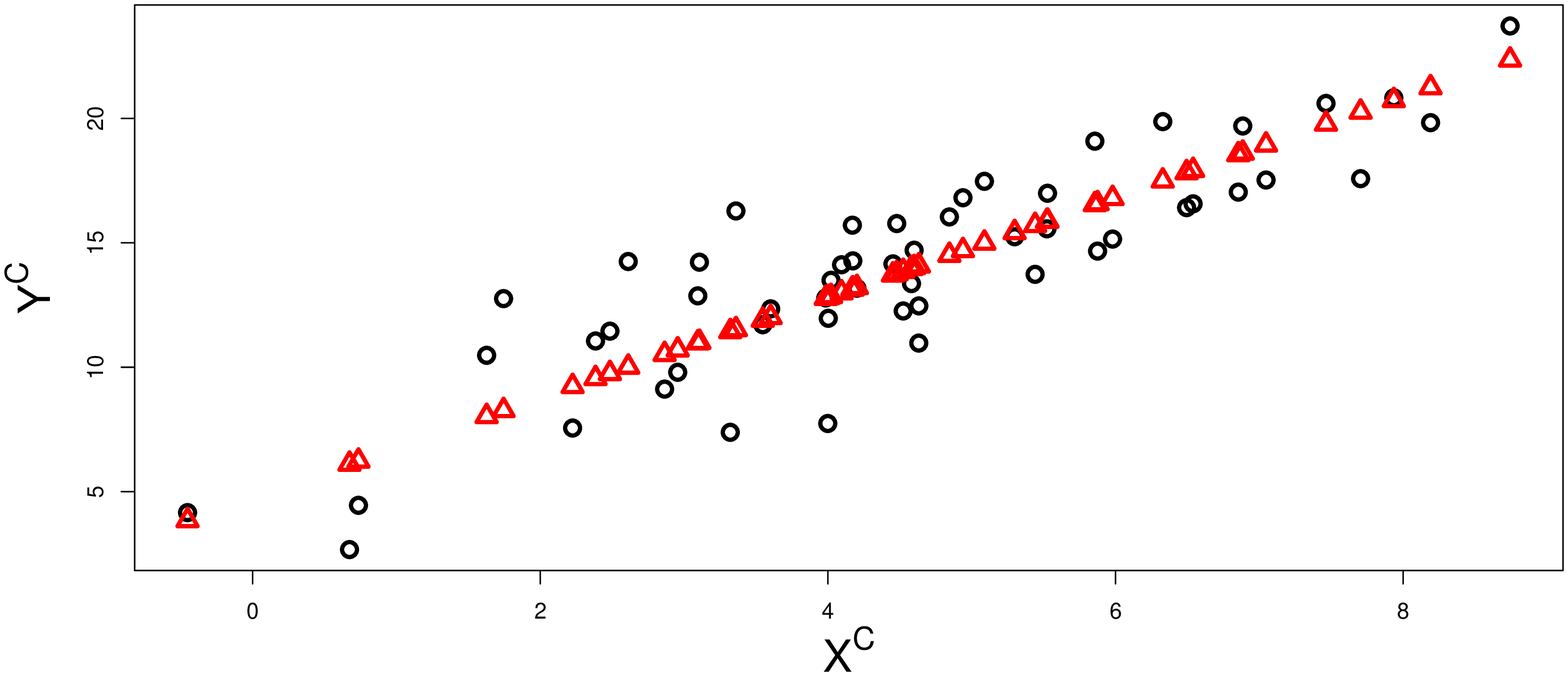}\\ 
	\includegraphics[width=170pt, height=110pt]{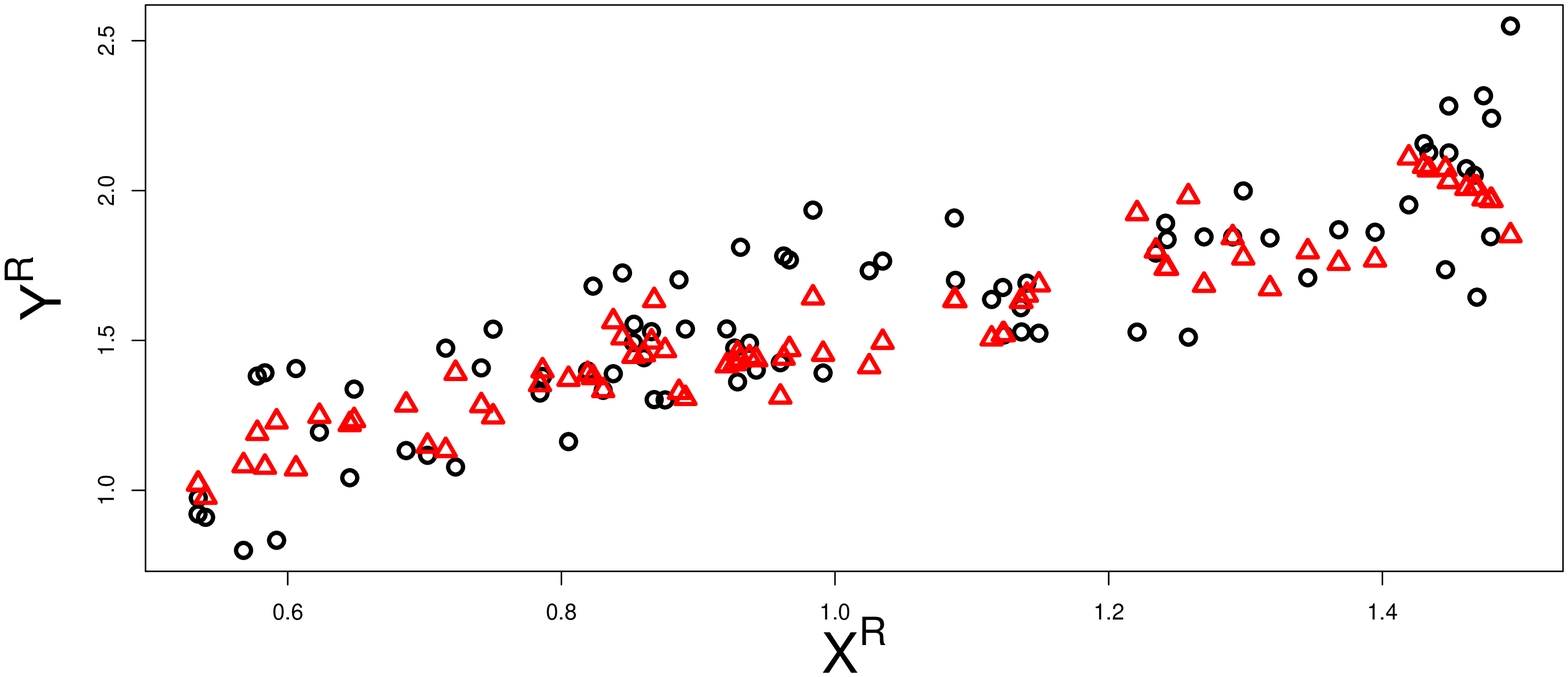} 
	\includegraphics[width=170pt, height=110pt]{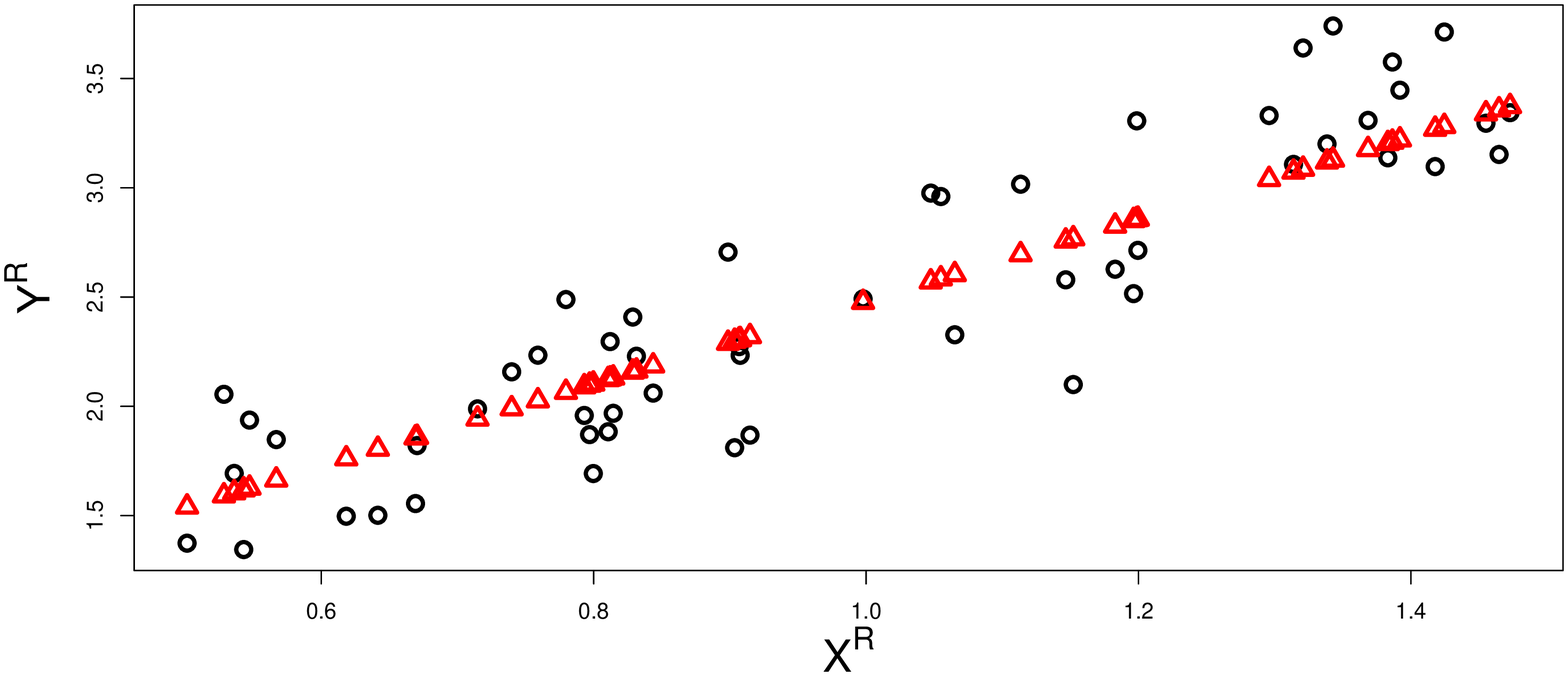}
	\caption{Overlaid plots of predicted centers and radii against the observed values from setting 1. The left two plots are for random forests and the right two are for CCRM. Observed data from the test set are represented by black circles and the predicted data are represented by red triangles (same for figures \ref{set2_pred}-\ref{set6_pred} in the following).}
	\label{set1_pred}
\end{figure}

\begin{figure}
	\centering
	\includegraphics[width=170pt, height=110pt]{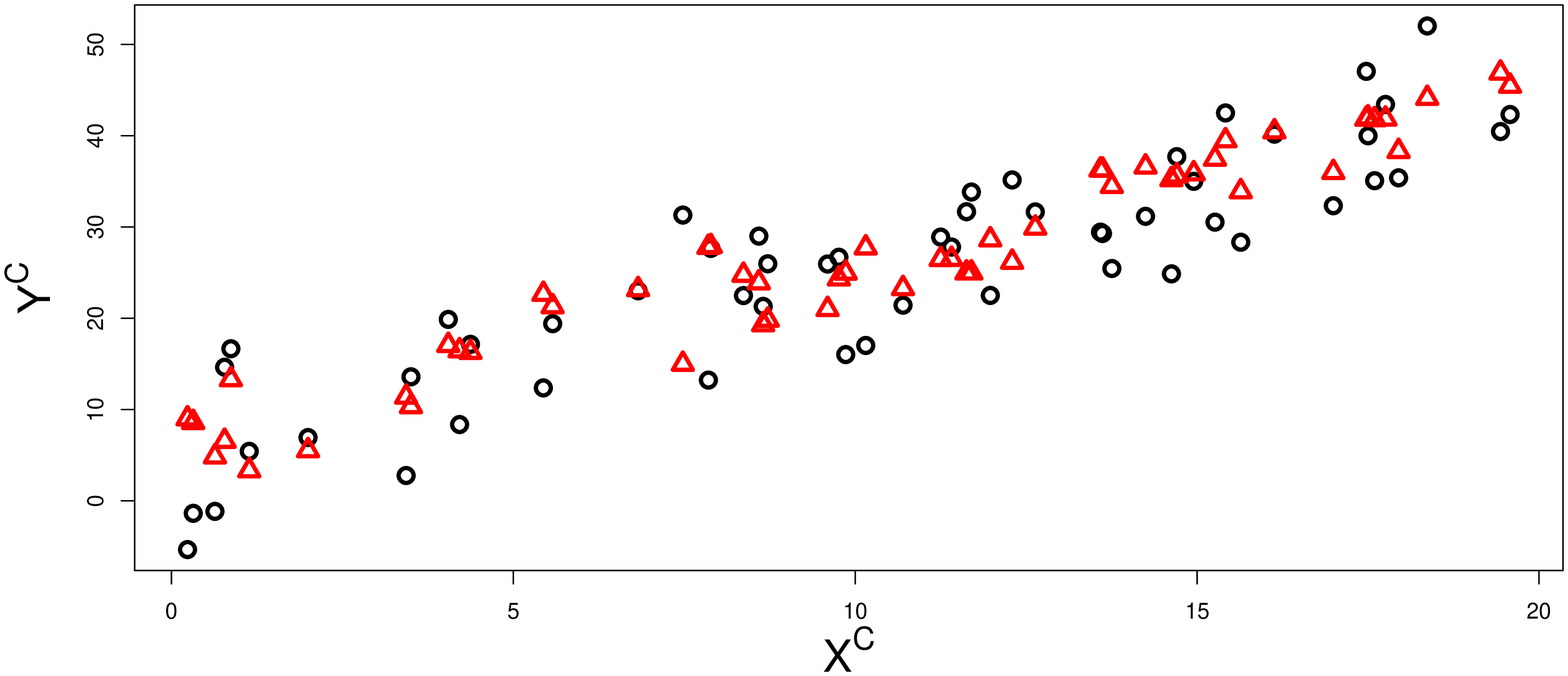} 
	\includegraphics[width=170pt, height=110pt]{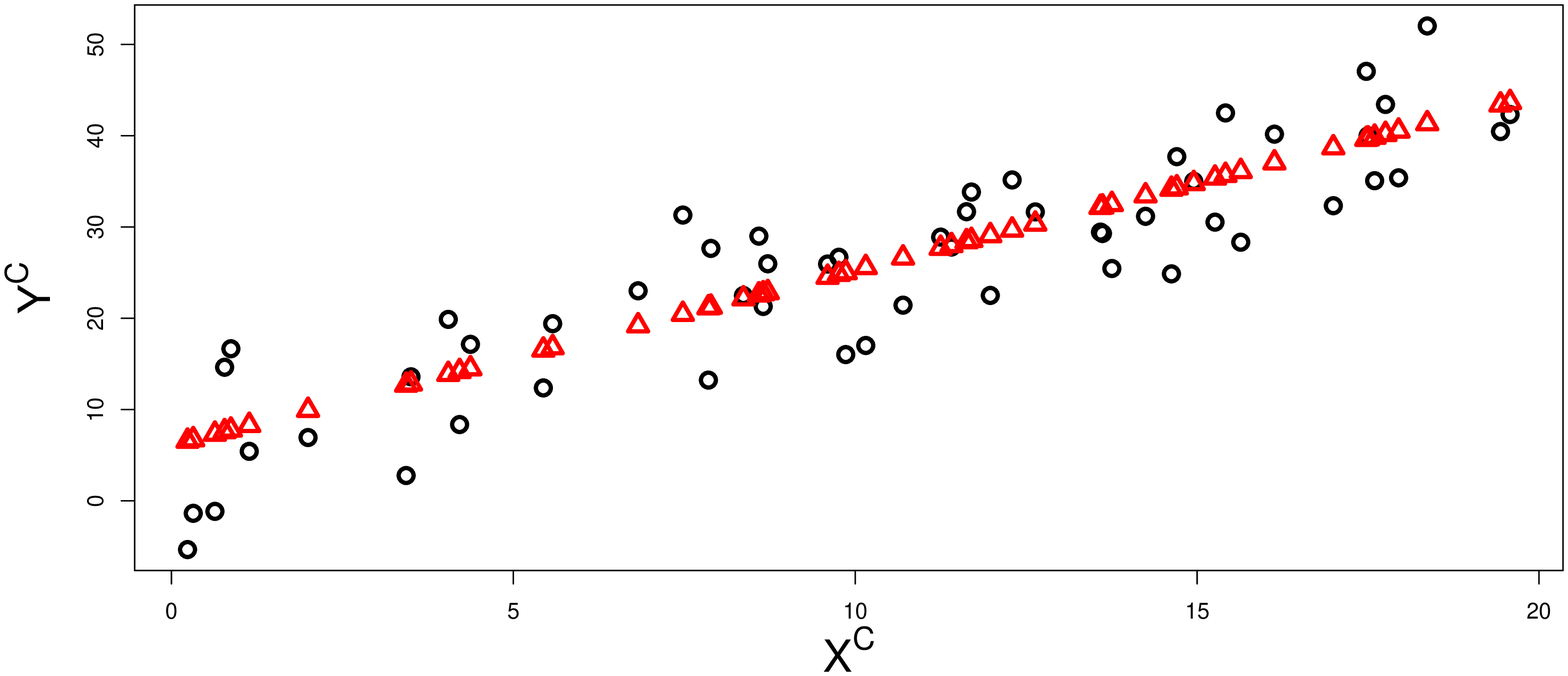}\\ 
	\includegraphics[width=170pt, height=110pt]{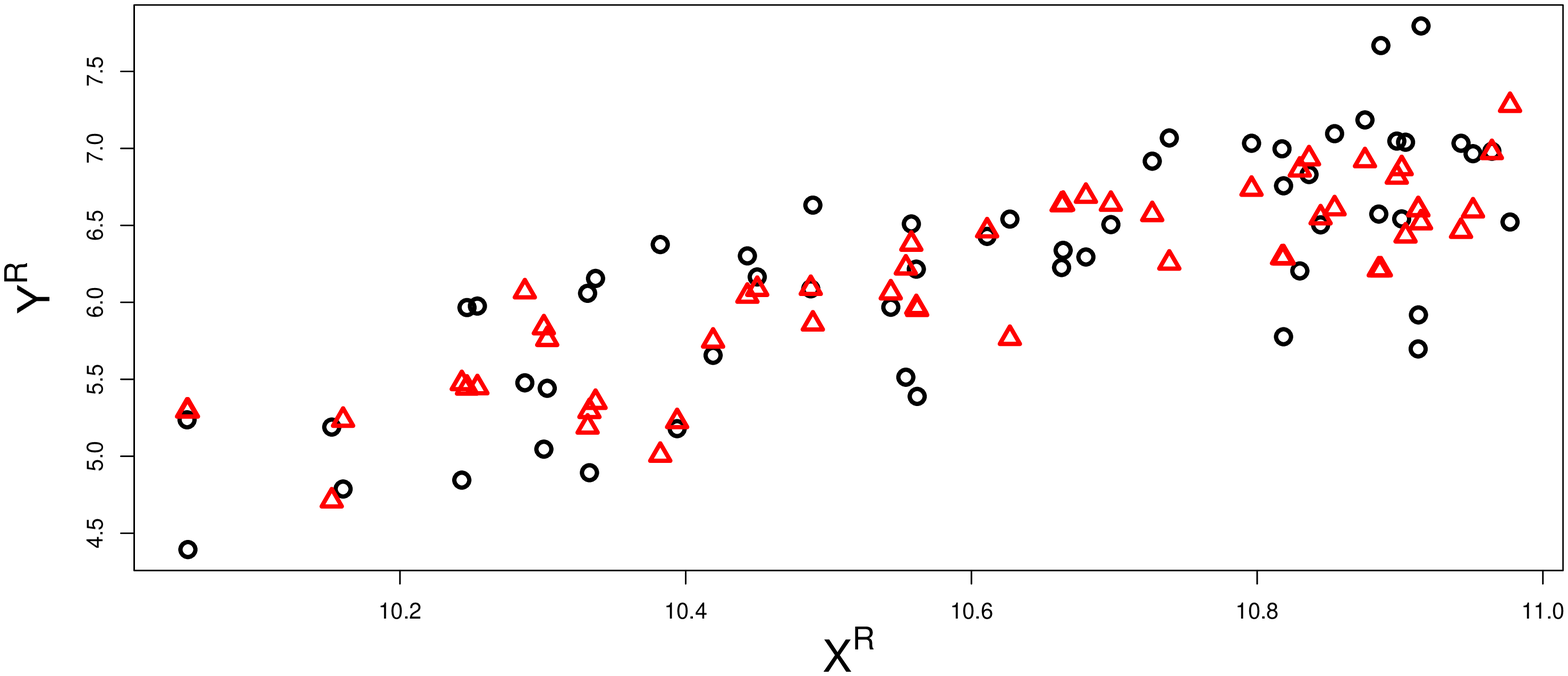} 
	\includegraphics[width=170pt, height=110pt]{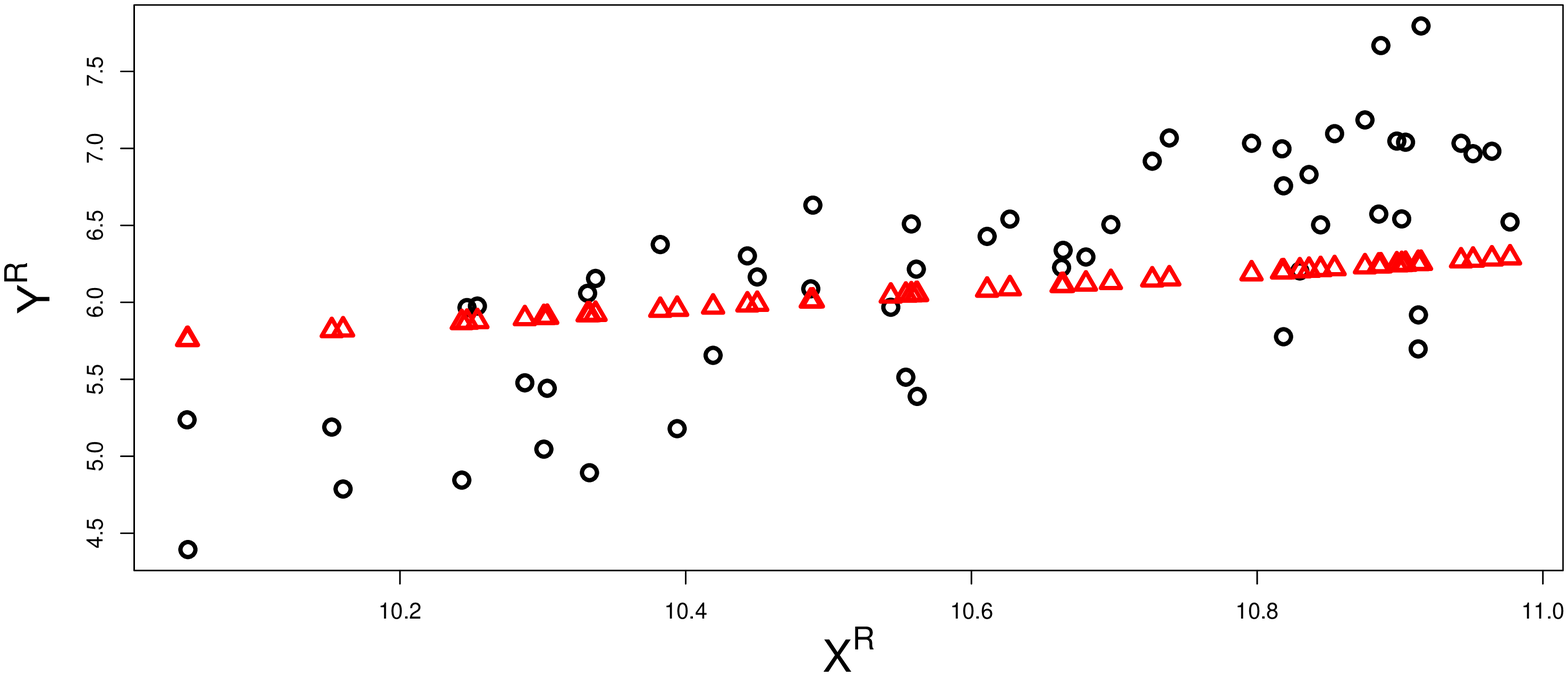}
	\caption{Overlaid plots of predicted centers and radii against the observed values from setting 2. The left two plots are for random forests and the right two are for CCRM.}
	\label{set2_pred}
\end{figure}

\begin{figure}
	\centering
	\includegraphics[width=170pt, height=110pt]{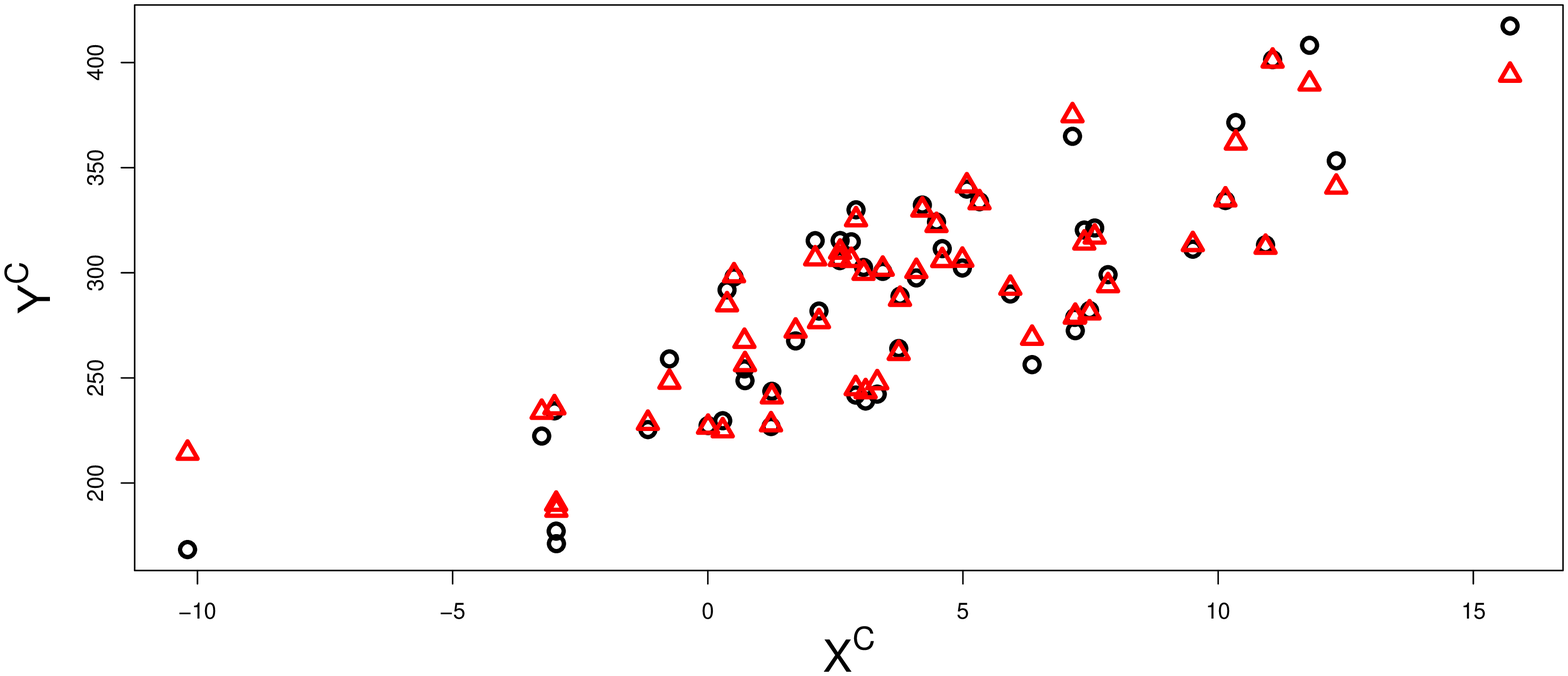} 
	\includegraphics[width=170pt, height=110pt]{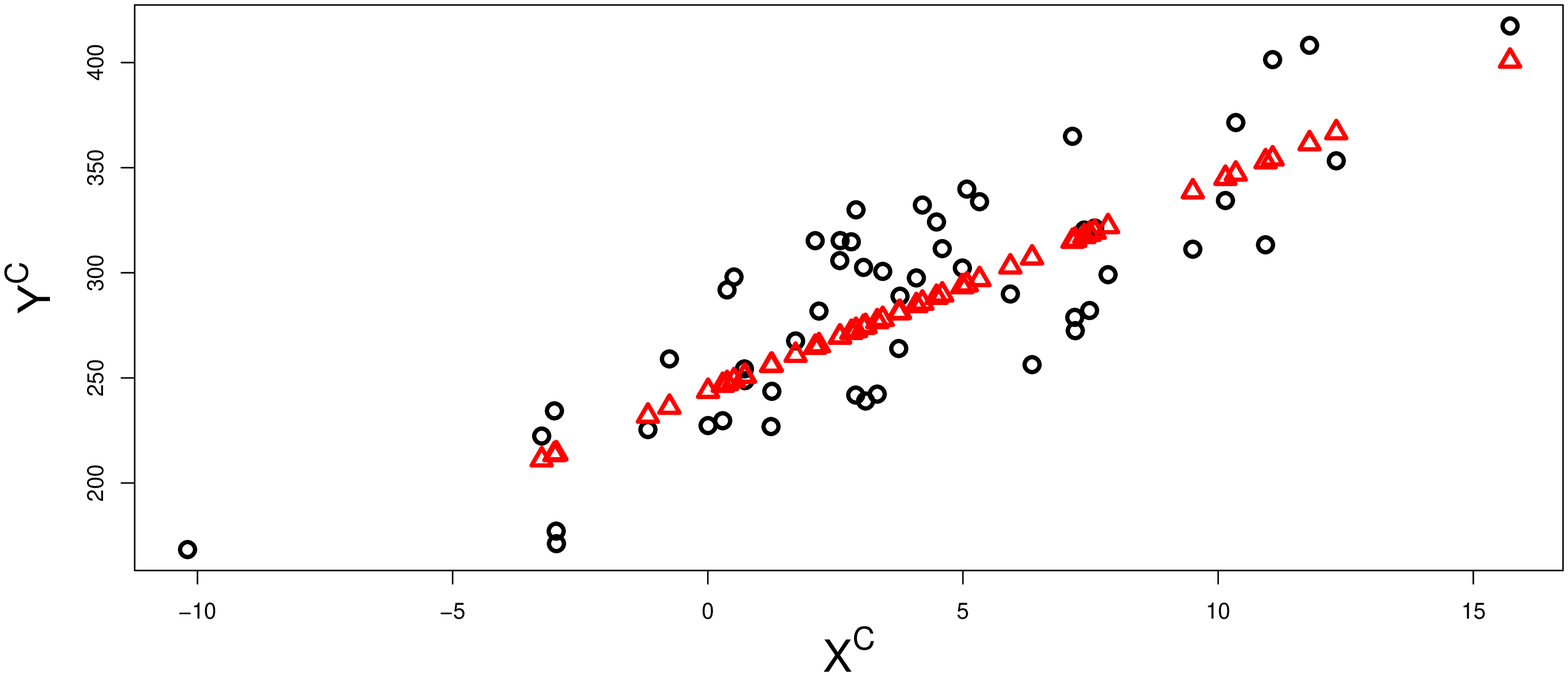}\\ 
	\includegraphics[width=170pt, height=110pt]{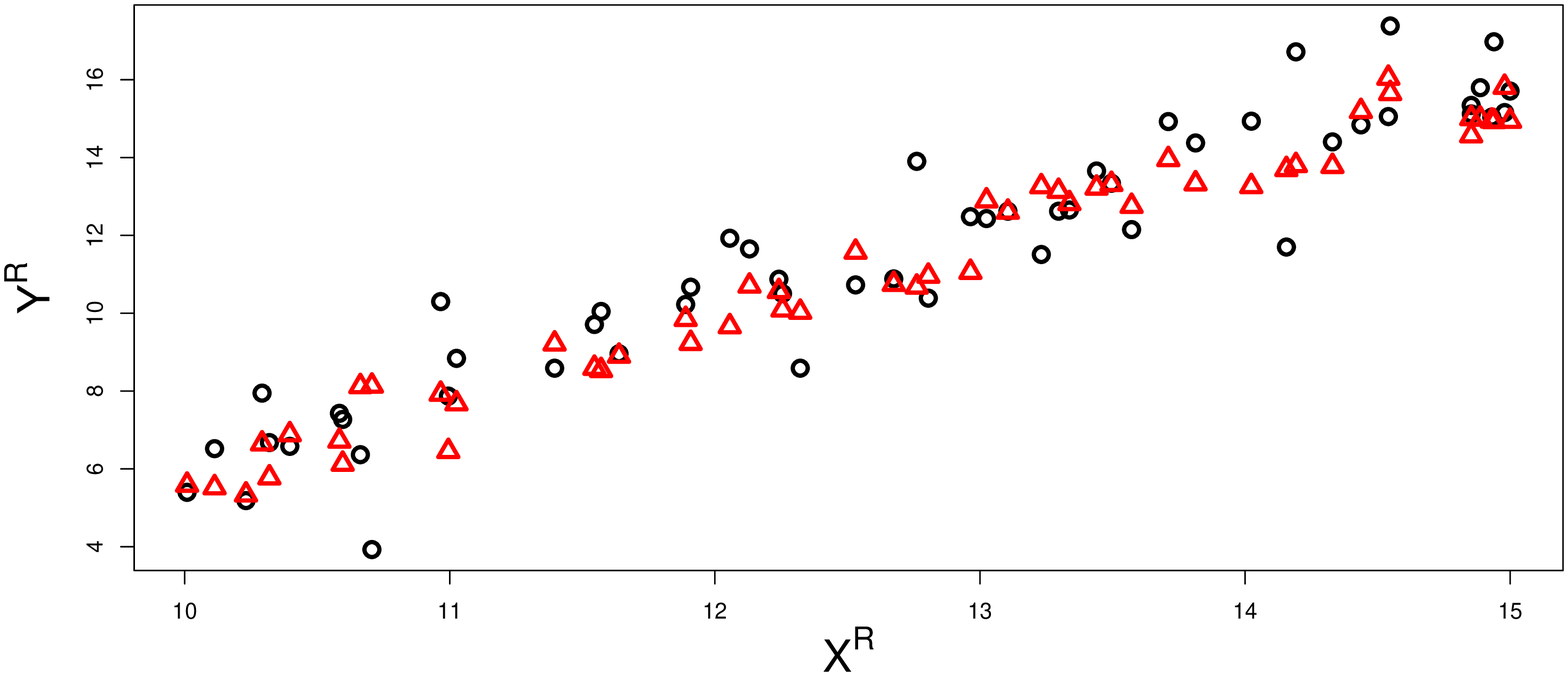} 
	\includegraphics[width=170pt, height=110pt]{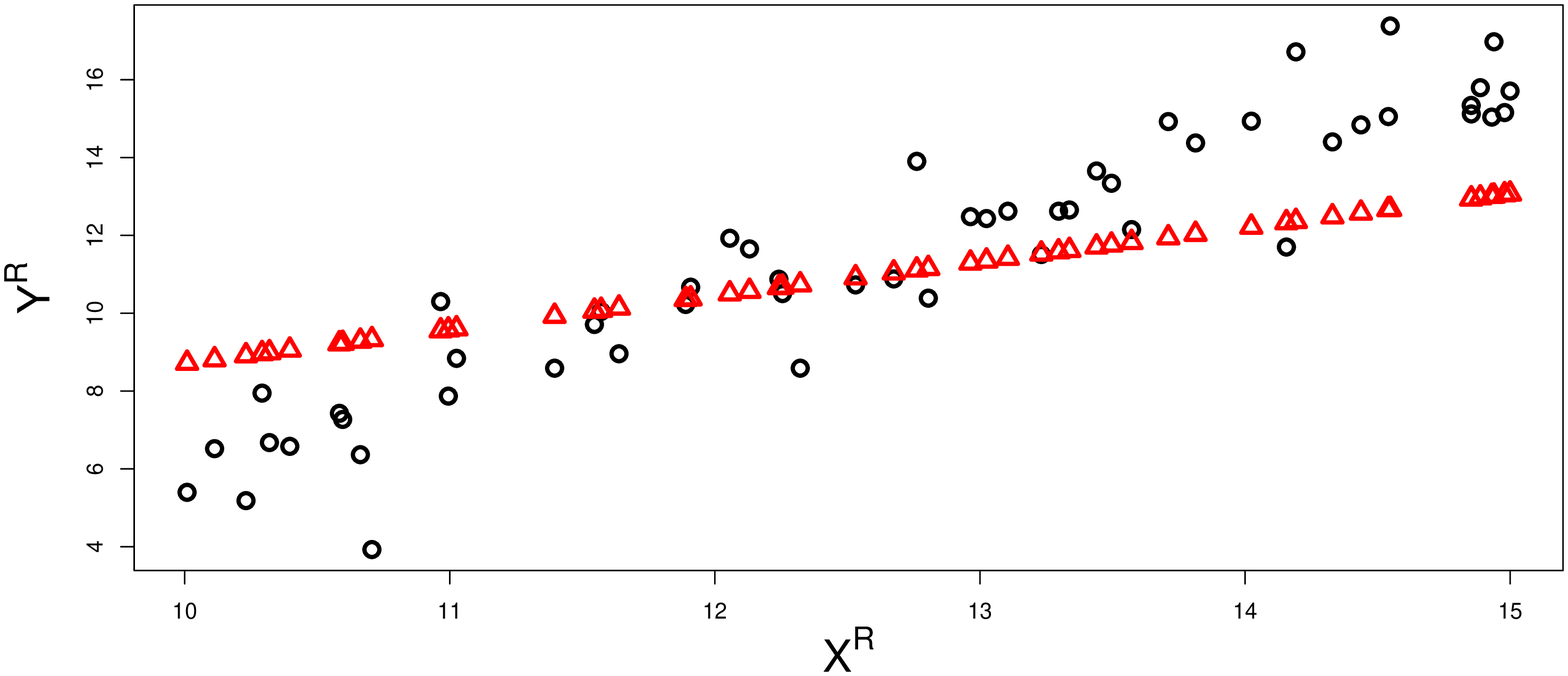}
	\caption{Overlaid plots of predicted centers and radii against the observed values from setting 3.The left two plots are for random forests and the right two are for CCRM.}
	\label{set3_pred}
\end{figure}

\begin{figure}
	\centering
	\includegraphics[width=170pt, height=110pt]{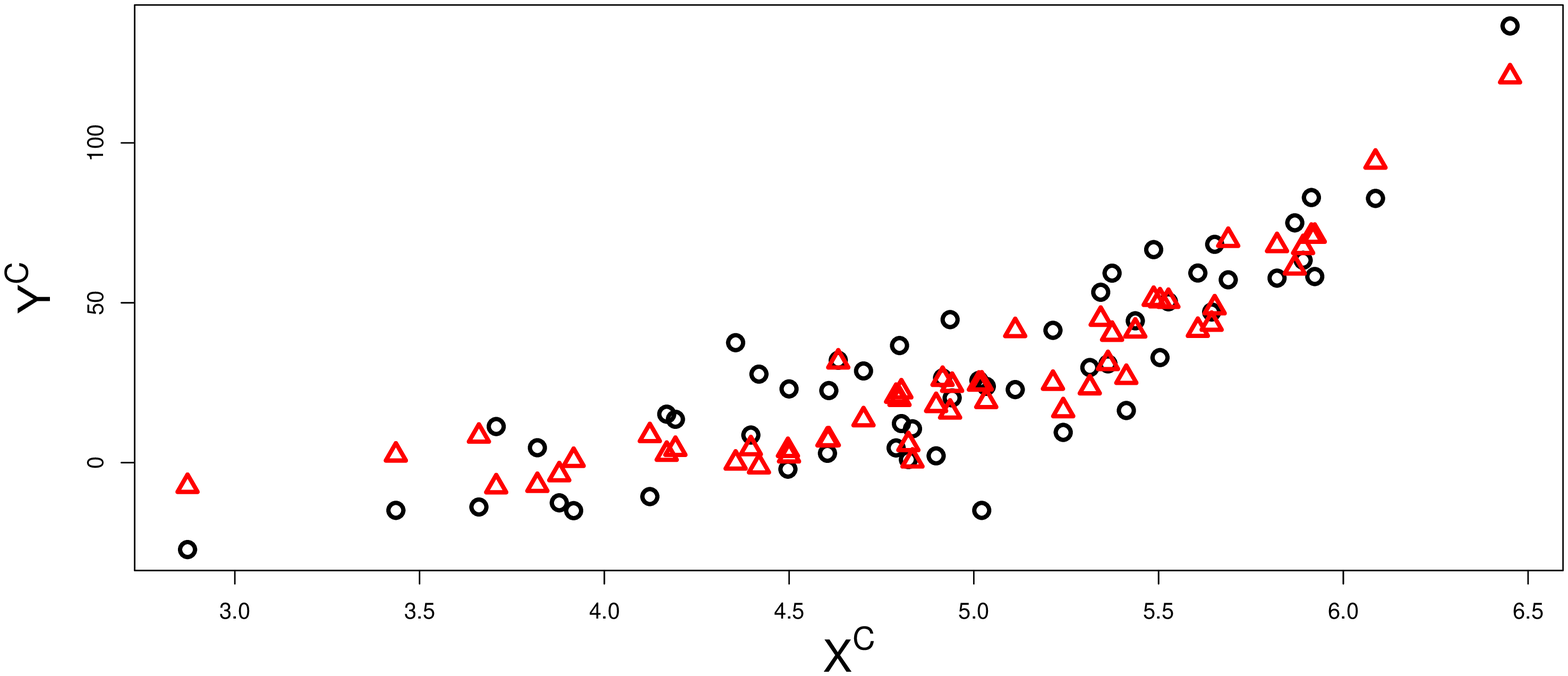} 
	\includegraphics[width=170pt, height=110pt]{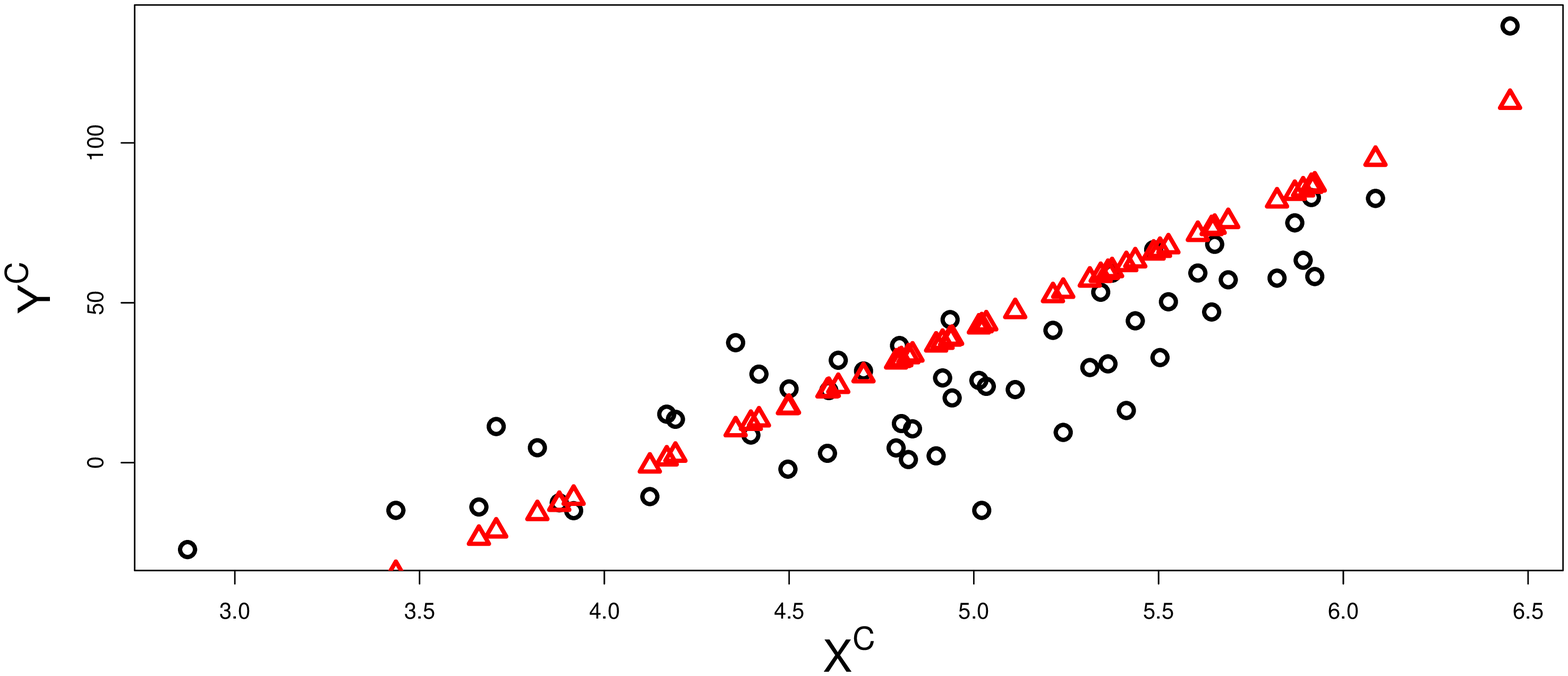}\\ 
	\includegraphics[width=170pt, height=110pt]{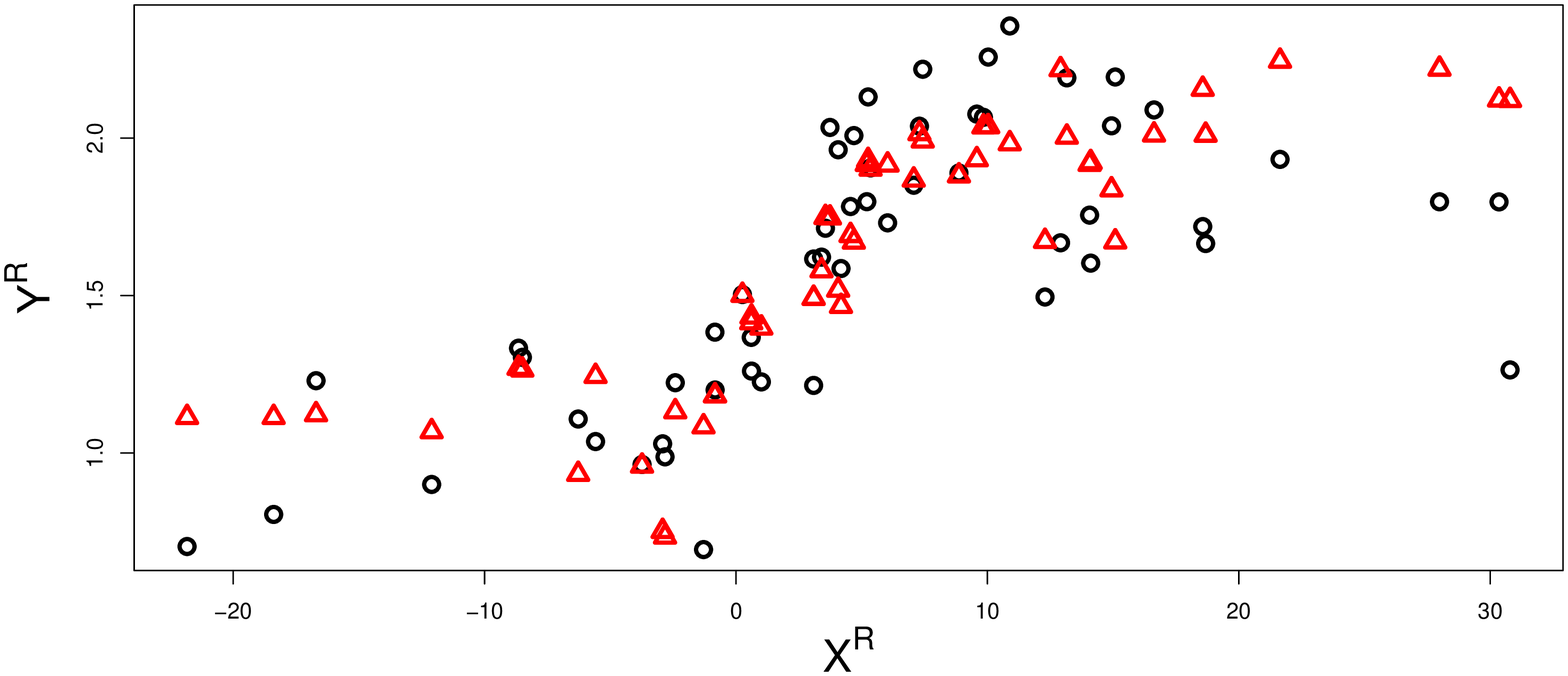} 
	\includegraphics[width=170pt, height=110pt]{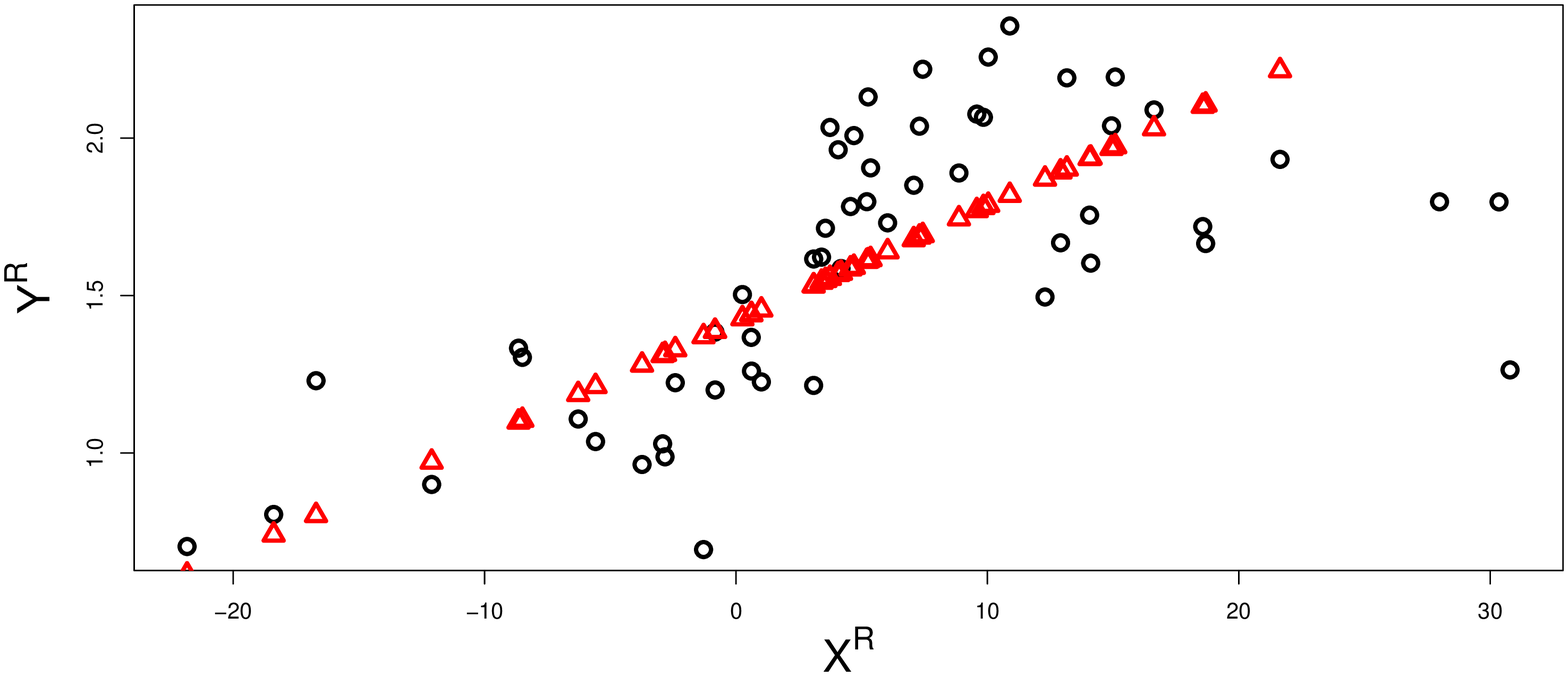}
	\caption{Overlaid plots of predicted centers and radii against the observed values from setting 4.The left two plots are for random forests and the right two are for CCRM.}
	\label{set4_pred}
\end{figure}

The predictive accuracies for the center of the linear models (settings 1-4) are displayed in table \ref{tab:LRc} and plotted in figures \ref{set1_pred} - \ref{set4_pred}. The performances of both CCRM and random forests generally improve for all of the four settings as the sample size increases. In the first two settings where the data show a strong linear association, random forest regression tree does not have any advantages over CCRM, - in fact, CCRM performs slightly better than random forests for these two settings. In setting 3, the response center is linearly associated with both the predictor center and radius, which CCRM is incapable of capturing but random forests can. Therefore, in this setting, random forest regression yields higher predictive accuracy over CCRM for the center. For the slightly nonlinear model in setting 4, the random forest regression outcompetes CCRM, due to its ability to handle nonlinearity.

\begin{table}[htbp]
	\centering
	\caption{Predictive Accuracy for Linear Models: Center (Settings 1-4)}
	\begin{tabular}{rrrrrrrr}
		\toprule
		&       & \multicolumn{2}{c}{$R^2$} & \multicolumn{2}{c}{MSE} & \multicolumn{2}{c}{MAE} \\
		\cmidrule{2-8}    \multicolumn{1}{c}{Setting} & \multicolumn{1}{c}{n} & \multicolumn{1}{l}{CCRM} & \multicolumn{1}{l}{RF} & \multicolumn{1}{l}{CCRM} & \multicolumn{1}{l}{RF} & \multicolumn{1}{l}{CCRM} & \multicolumn{1}{l}{RF} \\
		\midrule
		1     & 50    & \textbf{0.7928} & 0.7161 & \textbf{4.110} & 5.650  & \textbf{1.620} & 1.890 \\
		& 100   & \textbf{0.7954} & 0.7259 & \textbf{4.050} & 5.450  & \textbf{1.610} & 1.860 \\
		& 200   & \textbf{0.7977} & 0.7278 & \textbf{4.040} & 5.430  & \textbf{1.600} & 1.860 \\
		&       &       &       &       &       &       &  \\
		2     & 50    & \textbf{0.8349} & 0.7868 & \textbf{26.08} & 33.62 & \textbf{4.080} & 4.630 \\
		& 100   & \textbf{0.8395} & 0.7878 & \textbf{25.26} & 33.45 & \textbf{4.010} & 4.630 \\
		& 200   & \textbf{0.8410} & 0.7894 & \textbf{25.16} & 33.31 & \textbf{4.000} & 4.610 \\
		&       &       &       &       &       &       &  \\
		3     & 50    & 0.7309 & \textbf{0.8662} & 883.6 & \textbf{446.4} & 25.53 & \textbf{14.81} \\
		& 100   & 0.7385 & \textbf{0.9276} & 866.7 & \textbf{241.6} & 25.35 & \textbf{10.55} \\
		& 200   & 0.7449 & \textbf{0.9599} & 854.6 & \textbf{133.9} & 25.22 & \textbf{7.610} \\
		&       &       &       &       &       &       &  \\
		4     & 50    & 0.5606 & \textbf{0.7157} & 1468.4 & \textbf{981.9} & 24.01 & \textbf{18.46} \\
		& 100   & 0.5447 & \textbf{0.7650} & 1474.1 & \textbf{834.4} & 24.38 & \textbf{17.50} \\
		& 200   & 0.5798 & \textbf{0.7973} & 1383.7 & \textbf{661.6} & 23.52 & \textbf{17.29} \\
		\bottomrule
	\end{tabular}%
	\label{tab:LRc}%
\end{table}%

Table \ref{tab:LRr} shows the predictive accuracies for the radius of the linear models. As for the center, the random forest regression is slightly worse than CCRM in setting 1 where the data follow a linear pattern with a strictly positive intercept. However, for linear models with a negative intercept in setting 2 and 3, CCRM returns poor predictive accuracy due to its constraint that the intercept must be positive, but random forests need not have this constraint and therefore perform much better. Also for the slightly nonlinear model in setting 4, random forests yield much better predictive accuracy than CCRM.
\leavevmode

\begin{table}[htbp]
	\centering
	\caption{Predictive Accuracy for Linear Models: Radius (Settings 1-4)}
	\begin{tabular}{rrrrrrrr}
		\toprule
		&       & \multicolumn{2}{c}{$R^2$} & \multicolumn{2}{c}{MSE} & \multicolumn{2}{c}{MAE} \\
		\cmidrule{3-8}    \multicolumn{1}{c}{Setting } & \multicolumn{1}{c}{n} & \multicolumn{1}{l}{CCRM} & \multicolumn{1}{l}{RF} & \multicolumn{1}{l}{CCRM} & \multicolumn{1}{l}{RF} & \multicolumn{1}{l}{CCRM} & \multicolumn{1}{l}{RF} \\
		\midrule
		1     & 50    & \textbf{0.7768} & 0.7095 & \textbf{0.0900} & 0.1200  & \textbf{0.2400} & 0.2800 \\
		& 100   & \textbf{0.7829} & 0.7139 & \textbf{0.0900} & 0.1200  & \textbf{0.2400} & 0.2800 \\
		& 200   & \textbf{0.7853} & 0.7131 & \textbf{0.0900} & 0.1200  & \textbf{0.2400} & 0.2800 \\
		&       &       &       &       &       &       &  \\
		2     & 50    & 0.2671 & \textbf{0.4144} & 0.4300  & \textbf{0.3400} & 0.5300  & \textbf{0.4700} \\
		& 100   & 0.2720 & \textbf{0.4182} & 0.4300  & \textbf{0.3400} & 0.5300  & \textbf{0.4700} \\
		& 200   & 0.2766 & \textbf{0.4178} & 0.4200  & \textbf{0.3400} & 0.5200  & \textbf{0.4600} \\
		&       &       &       &       &       &       &  \\
		3     & 50    & 0.5901 & \textbf{0.8126} & 3.910  & \textbf{1.790} & 1.630  & \textbf{1.070} \\
		& 100   & 0.5905 & \textbf{0.8139} & 3.970  & \textbf{1.790} & 1.640  & \textbf{1.070} \\
		& 200   & 0.5940 & \textbf{0.8144} & 3.930  & \textbf{1.790} & 1.630  & \textbf{1.070} \\
		&       &       &       &       &       &       &  \\
		4     & 50    & 0.5685 & \textbf{0.7434} & 0.1100  & \textbf{0.0700} & 0.2700  & \textbf{0.1900} \\
		& 100   & 0.5982 & \textbf{0.7810} & 0.1000   & \textbf{0.0600} & 0.2500  & \textbf{0.1700} \\
		& 200   & 0.6098 & \textbf{0.7970} & 0.0900  & \textbf{0.0500} & 0.2500  & \textbf{0.1600} \\
		\bottomrule
	\end{tabular}%
	\label{tab:LRr}%
\end{table}%

\begin{figure}
	\centering
	\includegraphics[width=170pt, height=110pt]{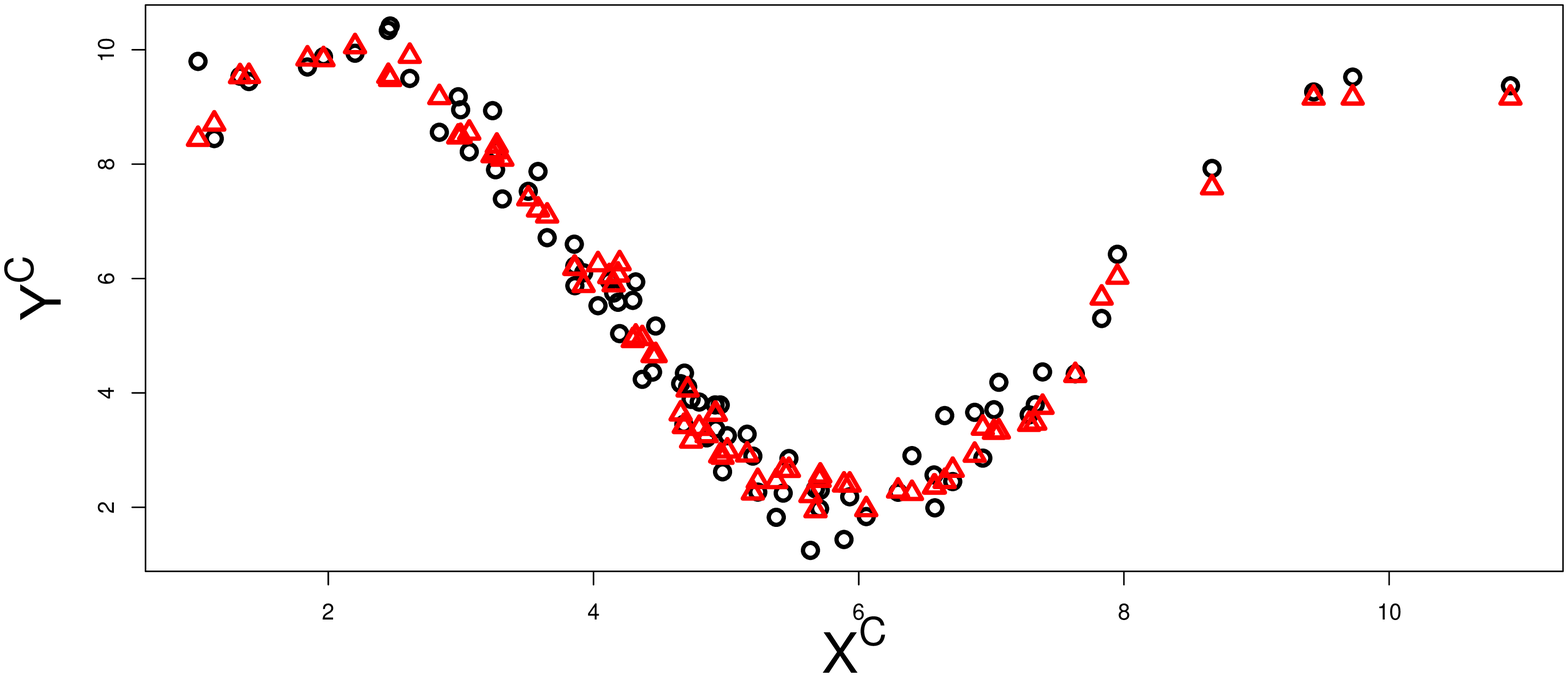} 
	\includegraphics[width=170pt, height=110pt]{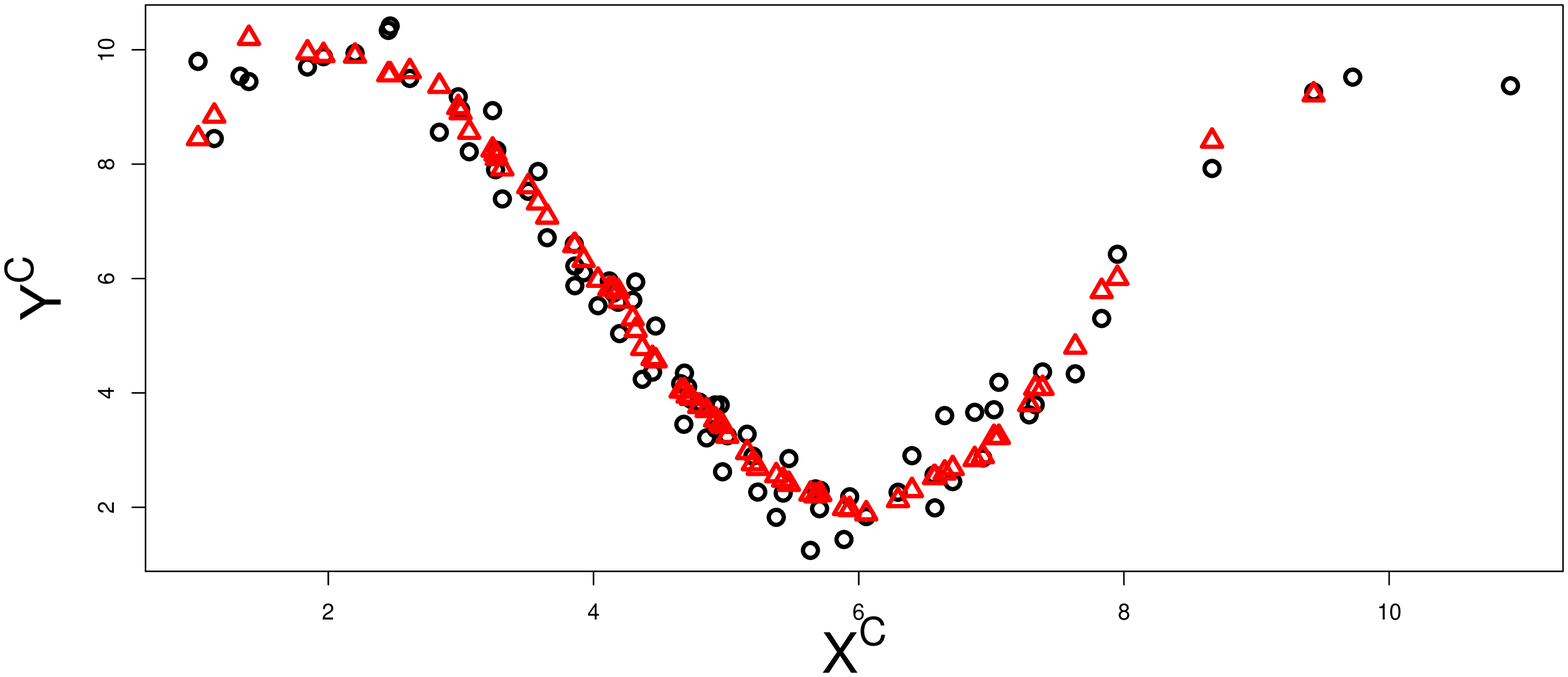}\\ 
	\includegraphics[width=170pt, height=110pt]{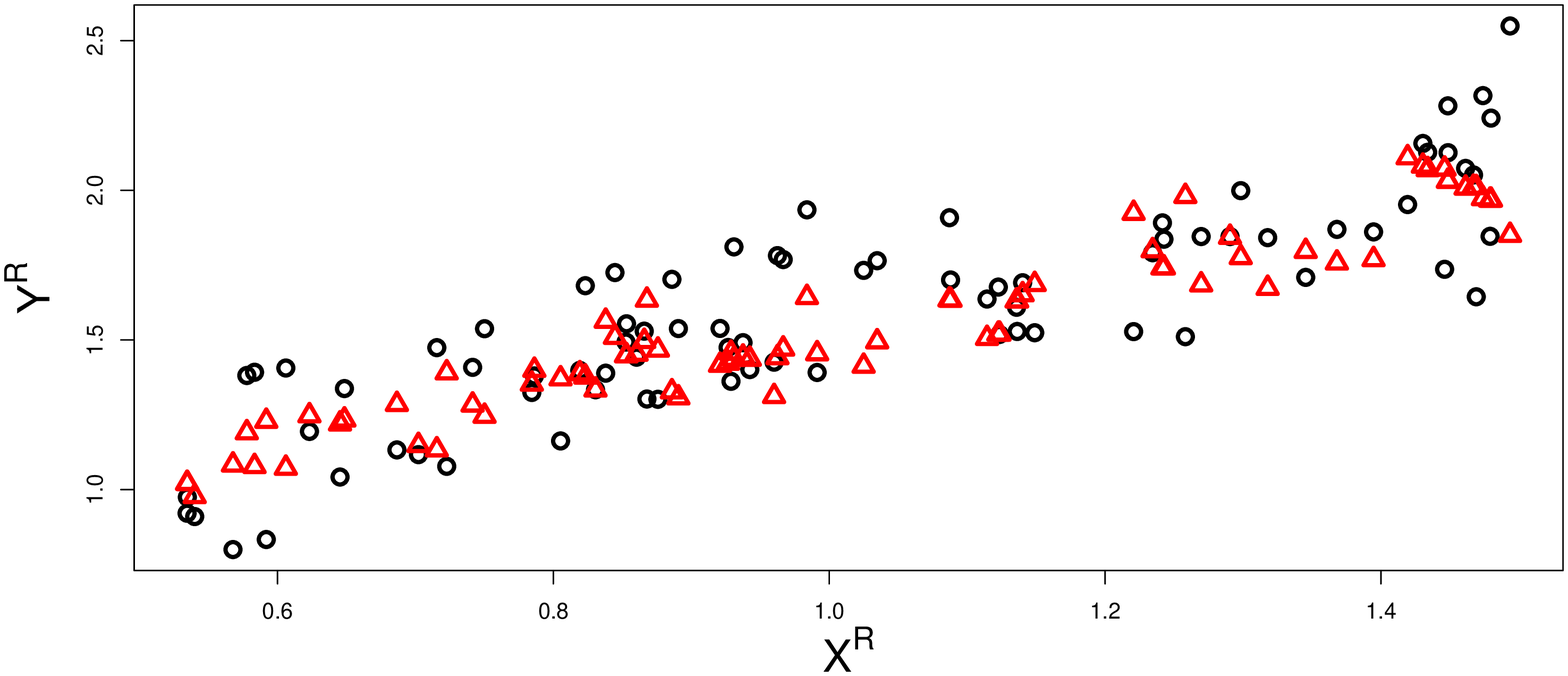} 
	\includegraphics[width=170pt, height=110pt]{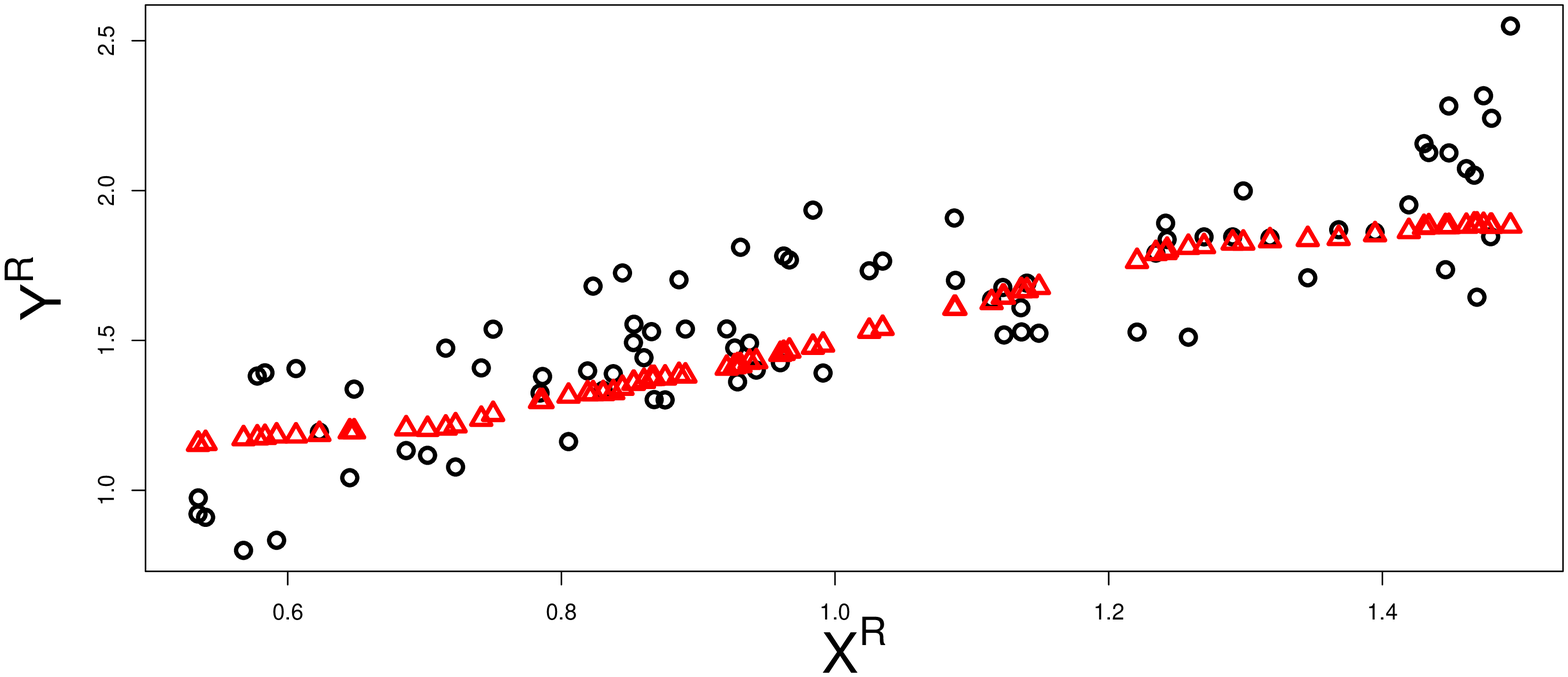}
	\caption{Overlaid plots of predicted centers and radii against the observed values from setting 5. The left two plots are for random forests and the right two are for kernel estimator.}
	\label{set5_pred}
\end{figure}

\begin{figure}
	\centering
	\includegraphics[width=170pt, height=110pt]{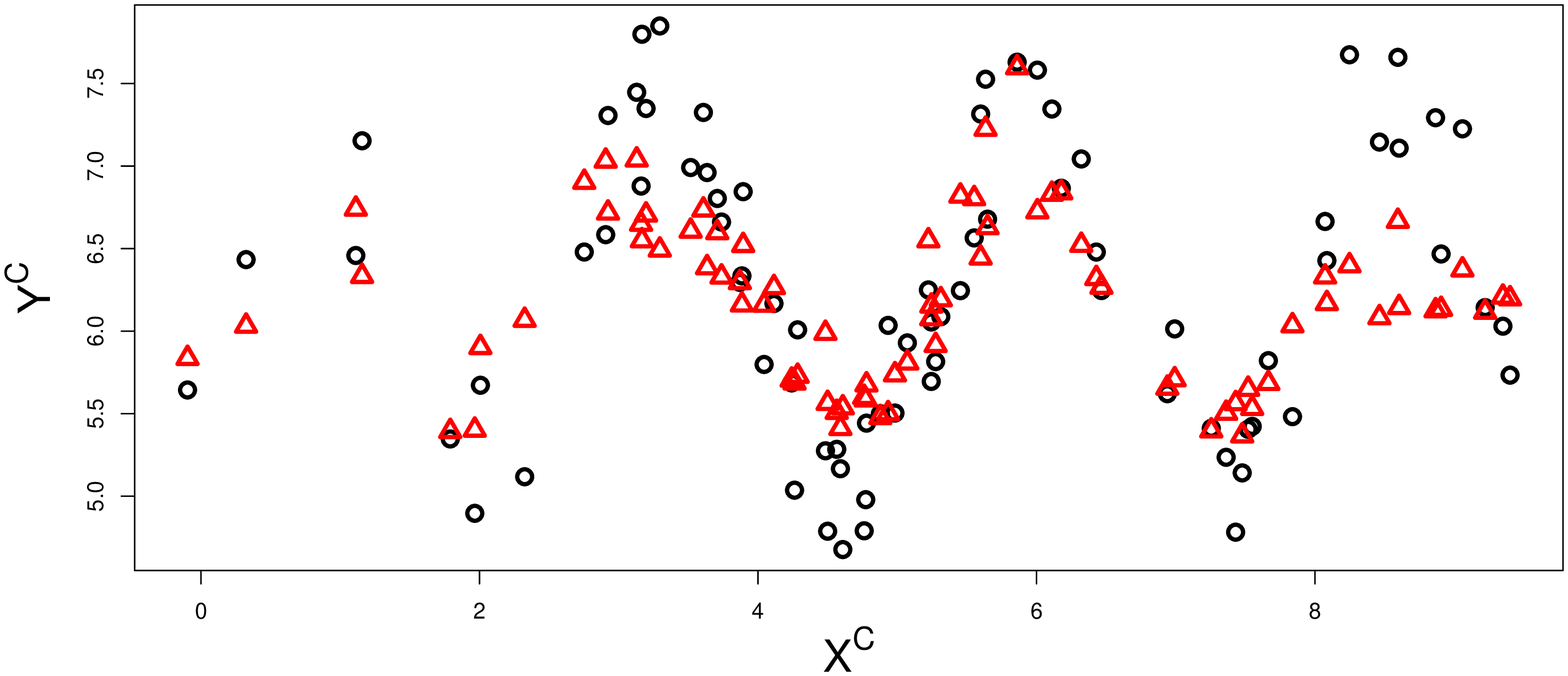} 
	\includegraphics[width=170pt, height=110pt]{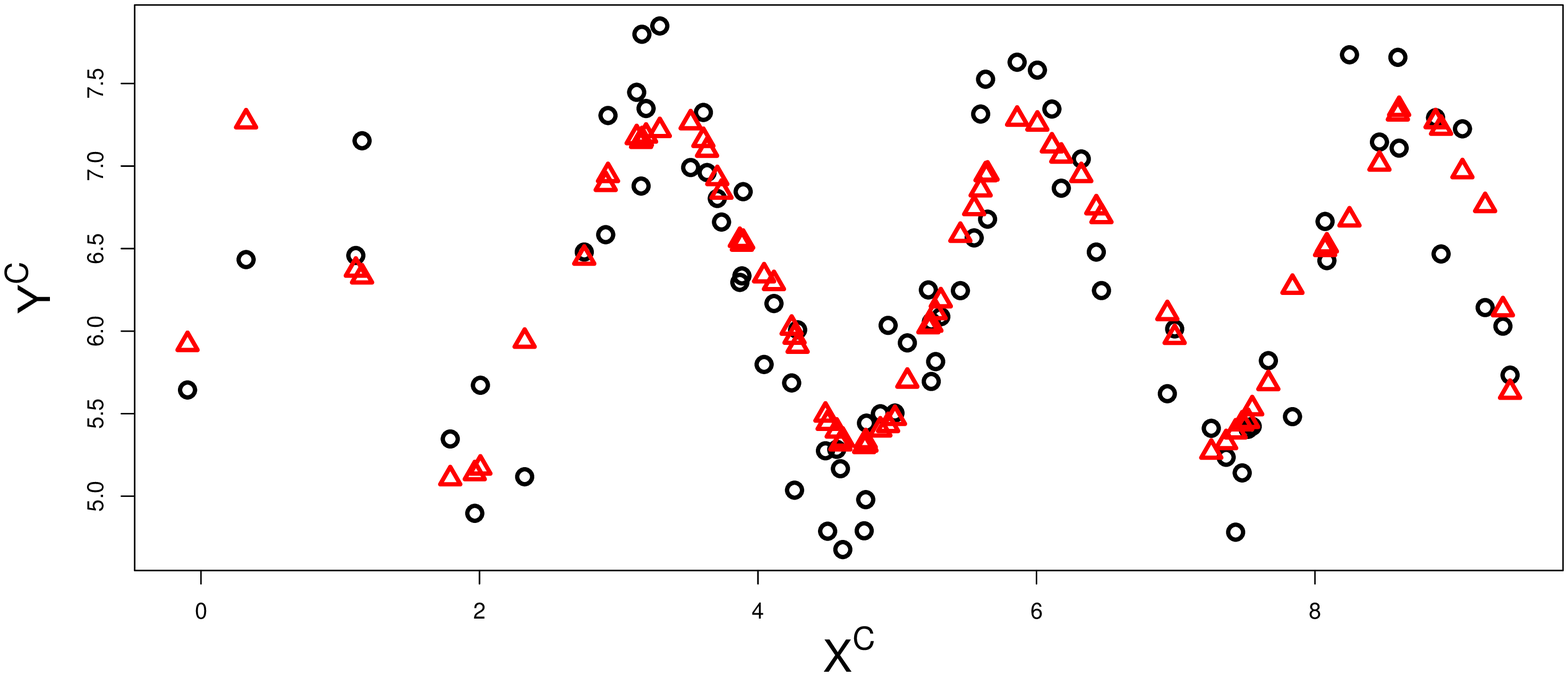}\\ 
	\includegraphics[width=170pt, height=110pt]{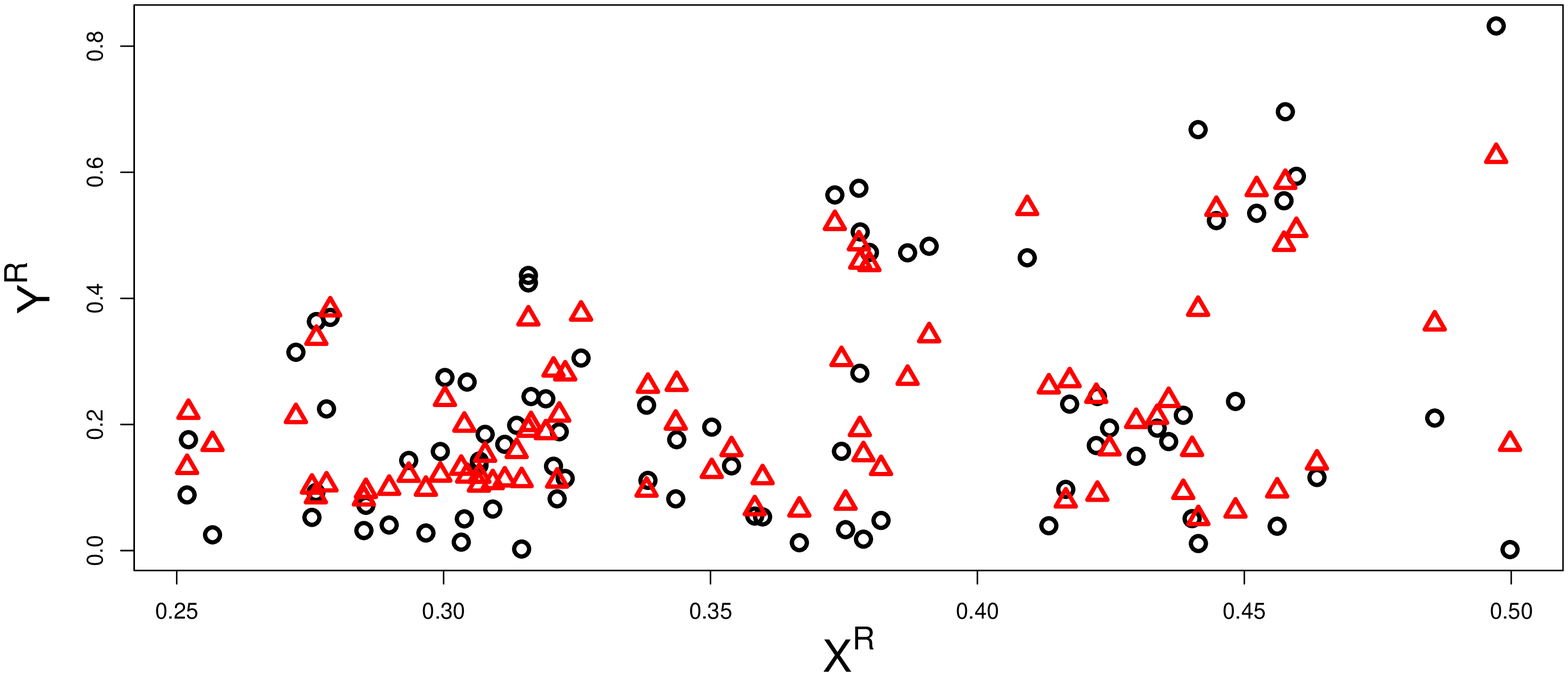} 
	\includegraphics[width=170pt, height=110pt]{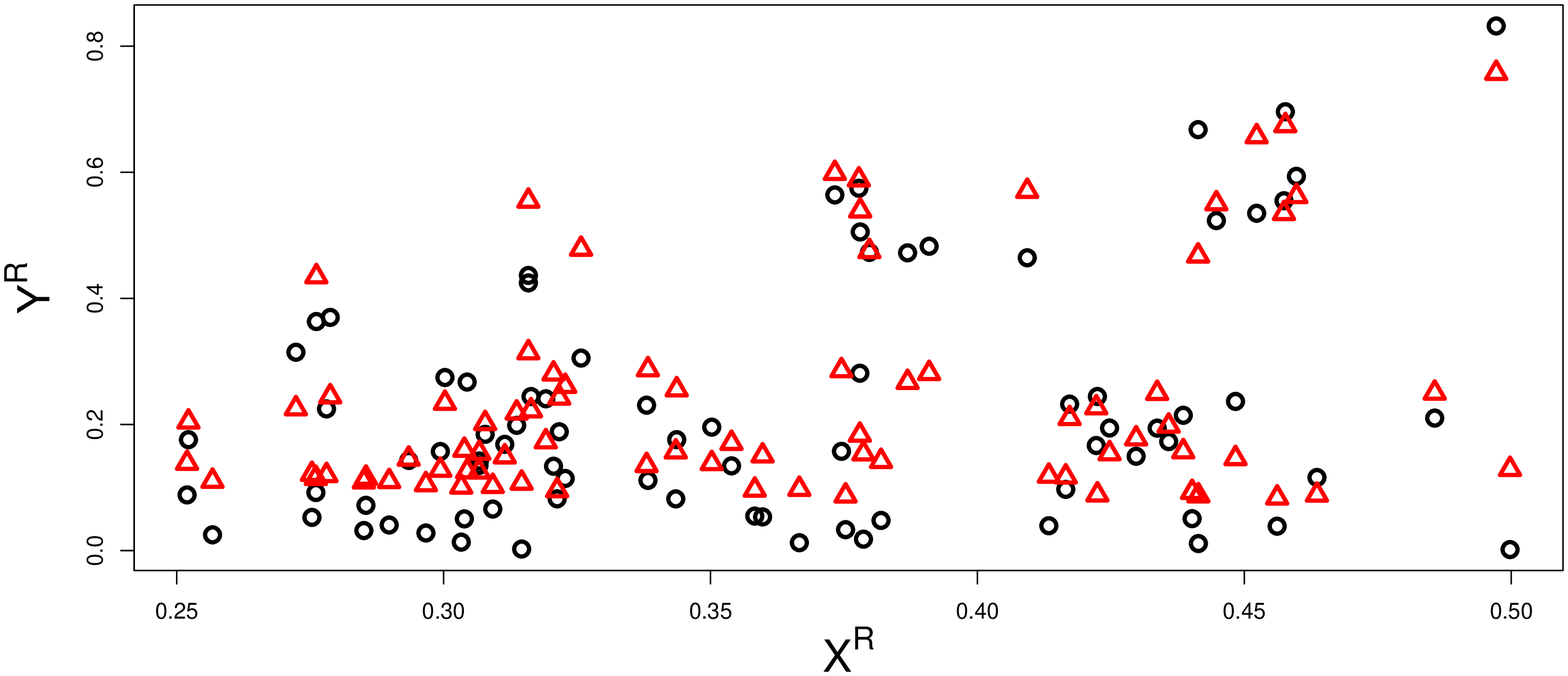}
	\caption{Overlaid plots of predicted centers and radii against the observed values from setting 6. The left two plots are for random forests and the right two are for kernel estimator.}
	\label{set6_pred}
\end{figure}

The center and radius results for nonlinear models are shown in table \ref{tab:NONc} and table \ref{tab:NONr}, respectively. The predictions against the original data are plotted in figures \ref{set5_pred} and \ref{set6_pred}. As for the linear models, as the sample size increases, both the kernel estimator and random forests improve with larger $R^{2}$s and smaller MSEs and MAEs. For the models with fewer predictors, such as setting 5 and 6, kernel estimator has competitive performance compared to random forest regression. For the multivariate model that has 5 predictors (setting 7), random forest regression has higher predictive accuracy than the kernel estimator, because the kernel-based estimator generally suffers from the ``curse of dimensionality''. Also, it is worth noting that, as the number of predictors rises, the kernel estimator rapidly becomes much more time consuming. For setting 7, the kernel estimator takes 28.12 seconds to run one simulation (for $n=50, 100, 200$ totally), while random forests only take 1.08 seconds.

\begin{table}[htbp]
  \centering
  \caption{Predictive Accuracy for Nonparametric Models: Center (Settings 5-7)}
    \begin{tabular}{rrrrrrrr}
    \toprule
          &       & \multicolumn{2}{c}{$R^2$} & \multicolumn{2}{c}{MSE} & \multicolumn{2}{c}{MAE} \\
\cmidrule{3-8}    \multicolumn{1}{c}{Setting } & \multicolumn{1}{c}{n} & \multicolumn{1}{c}{KE} & \multicolumn{1}{c}{RF} & \multicolumn{1}{c}{KE} & \multicolumn{1}{c}{RF} & \multicolumn{1}{c}{KE} & \multicolumn{1}{c}{RF} \\
    \midrule
5     & 50    & 0.6185 & \textbf{0.9030} & 2.880  & \textbf{0.7300} & 0.7800  & \textbf{0.5900} \\
          & 100   & 0.8288 & \textbf{0.9366} & 1.300   & \textbf{0.4800} & 0.5600  & \textbf{0.5100} \\
          & 200   & 0.9010 & \textbf{0.9480} & 0.7500  & \textbf{0.3900} & 0.4800  & \textbf{0.4800} \\
    6     & 50    & -0.0534 & \textbf{0.2644} & 0.8100  & \textbf{0.5600} & 0.7100  & \textbf{0.6000} \\
          & 100   & -0.2152 & \textbf{0.3872} & 0.9200  & \textbf{0.4700} & 0.6500  & \textbf{0.5400} \\
          & 200   & -0.0209 & \textbf{0.4746} & 0.7800  & \textbf{0.4000} & 0.5200  & \textbf{0.5000} \\
    7     & 50    & \textbf{0.5679} & 0.5276 & \textbf{17838} & 20370 & \textbf{63.76} & 99.59 \\
          & 100   & 0.6254 & \textbf{0.6403} & 15742 & \textbf{15380} & \textbf{55.82} & 84.88 \\
          & 200   & 0.6713 & \textbf{0.7340} & 13982 & \textbf{11410} & \textbf{47.19} & 71.20 \\
    \bottomrule
    \end{tabular}%
  \label{tab:NONc}%
\end{table}%

\begin{table}[htbp]
  \centering
  \caption{Predictive Accuracy for Nonlinear Models: Radius (Settings 5-7)}
    \begin{tabular}{rrrrrrrr}
    \toprule
          &       & \multicolumn{2}{c}{$R^2$} & \multicolumn{2}{c}{MSE} & \multicolumn{2}{c}{MAE} \\
\cmidrule{3-8}    \multicolumn{1}{c}{Setting } & \multicolumn{1}{c}{n } & \multicolumn{1}{c}{KE} & \multicolumn{1}{c}{RF} & \multicolumn{1}{c}{KE} & \multicolumn{1}{c}{RF} & \multicolumn{1}{c}{KE} & \multicolumn{1}{c}{RF} \\
    \midrule
    5     & 50    & \textbf{0.6341} & 0.5527 & \textbf{0.0400} & 0.0500  & \textbf{0.1700} & 0.1900 \\
          & 100   & \textbf{0.6497} & 0.5610 & \textbf{0.0400} & 0.0500  & \textbf{0.1700} & 0.1900 \\
          & 200   & \textbf{0.6608} & 0.5600  & \textbf{0.0400} & 0.0500  & \textbf{0.1600} & 0.1900 \\
    6     & 50    & 0.3425 & \textbf{0.4546} & 0.0200  & \textbf{0.0200} & 0.1000   & \textbf{0.1000} \\
          & 100   & 0.4764 & \textbf{0.5743} & 0.0200  & \textbf{0.0100} & 0.0900  & \textbf{0.0900} \\
          & 200   & 0.5419 & \textbf{0.6328} & 0.0100  & \textbf{0.0100} & 0.0800  & \textbf{0.0800} \\
    7     & 50    & 0.1321 & \textbf{0.6065} & 7.430  & \textbf{3.520} & \textbf{1.250} & 1.410 \\
          & 100   & 0.1561 & \textbf{0.7033} & 9.440  & \textbf{2.690} & 1.230  & \textbf{1.220} \\
          & 200   & 0.2326 & \textbf{0.7900} & 6.970  & \textbf{1.900} & \textbf{0.9600} & 1.000 \\
    \bottomrule
    \end{tabular}
  \label{tab:NONr}
\end{table}

\section{Real Data Application}\label{sec:application}

A real dataset is analyzed by the random forest regression to show its applicability. The dataset contains daily [min, max] stock price ranges for three companies, namely Boeing Aircraft Manufacturing Company (BA), General Electric (GE), and JPMorgan Chase (JPM), and the Down Jones Industrial Average index (DJIA), with 1511 price intervals for each asset from Jan 3, 2006 to Dec 30, 2011. We split the data into a training set of 1208 (80\%) and a test set of 303 (20\%) intervals.

\begin{figure} 
	\centering
	\includegraphics[width=350pt]{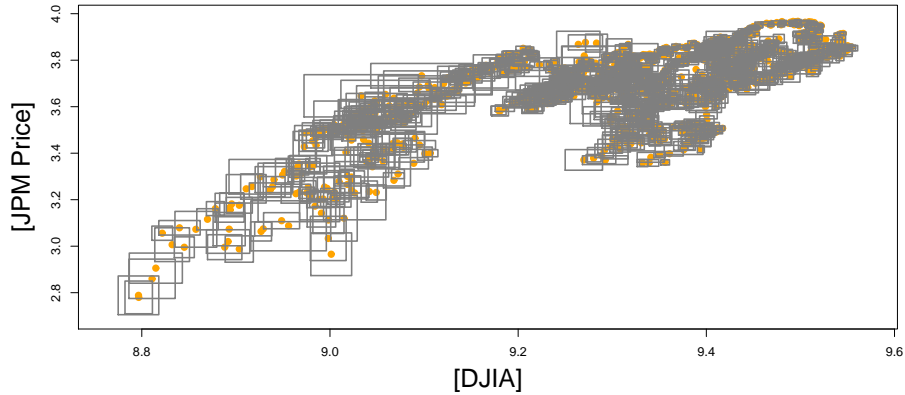}
	\caption{Plot of the Stock Price Interval for JPM. The gray rectangles denote the interval-valued data, and the yellow dots are the corresponding centers.}
	\label{DJ_JPM_big}
\end{figure}

DJIA is a stock market index created by Wall Street Journal editor and Dow Jones \& Company co-founder Charles Dow, to show how 30 large, publicly owned, companies based in the United States have traded during a standard trading session in the stock market. In our analysis, the DJIA is first used as the sole variable to predict each of the three individual stocks. As seen in figure \ref{DJ_JPM_big}, JPM (as well as other stocks) represents a pretty linear relationship with the DJIA index. So CCRM is first utilized as the baseline model. Then, random forest regressions are built for the centers and radii of the stocks, using both the center and radius of the DJIA index as predictors. Predictive accuracies are assessed in terms of the $R^2$, $MSE$ and $MAE$ on the test sets, and are reported in tables \ref{tab:center_table} and \ref{tab:range_table}. 

\begin{figure} 
	\centering
	\includegraphics[width=170pt, height=110pt]{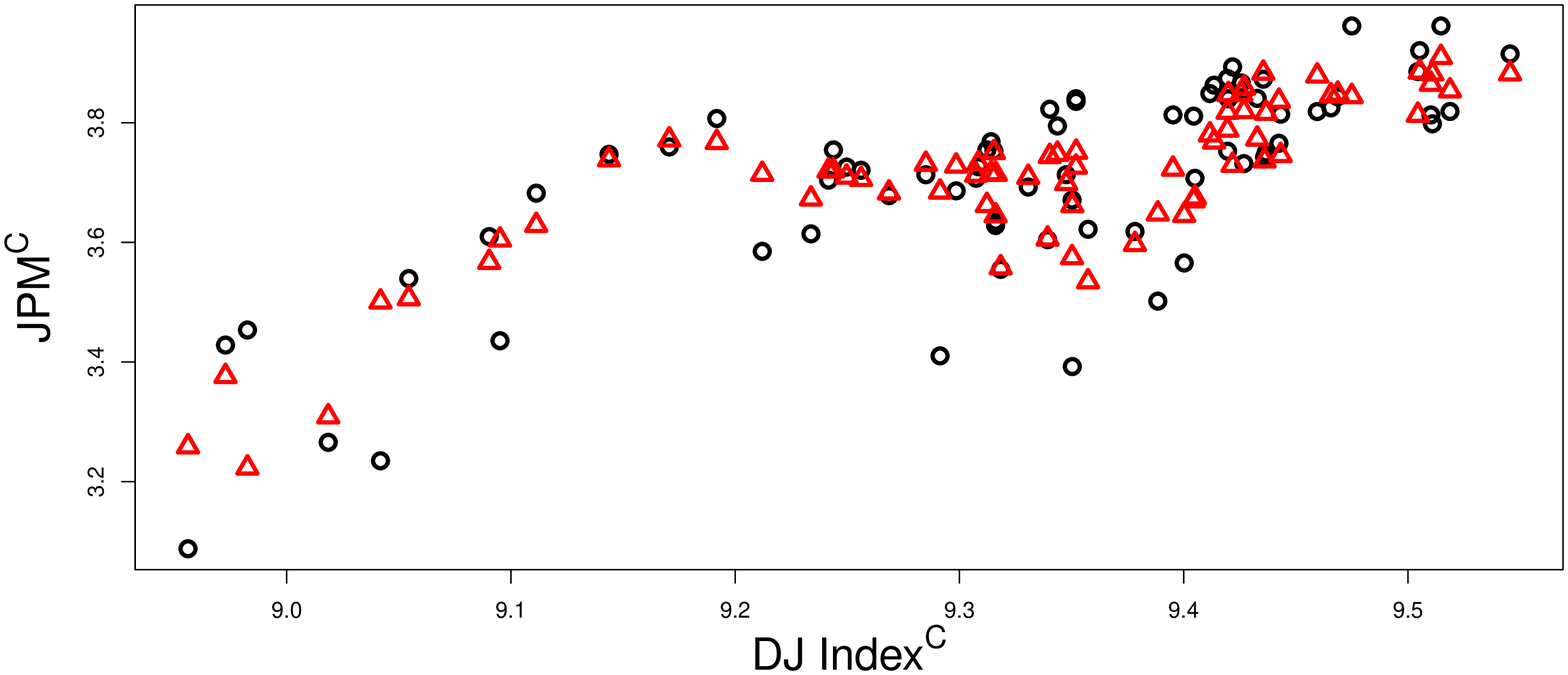} 
	\includegraphics[width=170pt, height=110pt]{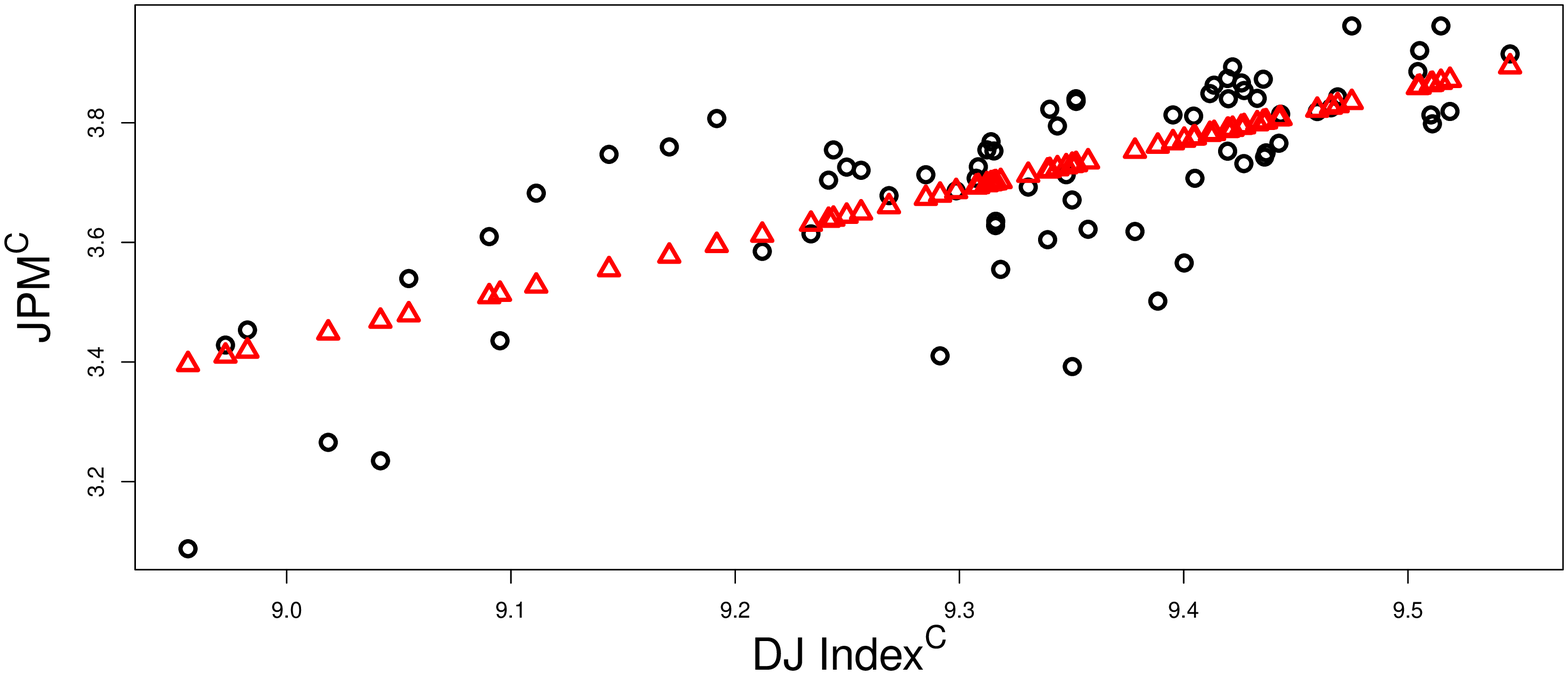}\\ 
	\includegraphics[width=170pt, height=110pt]{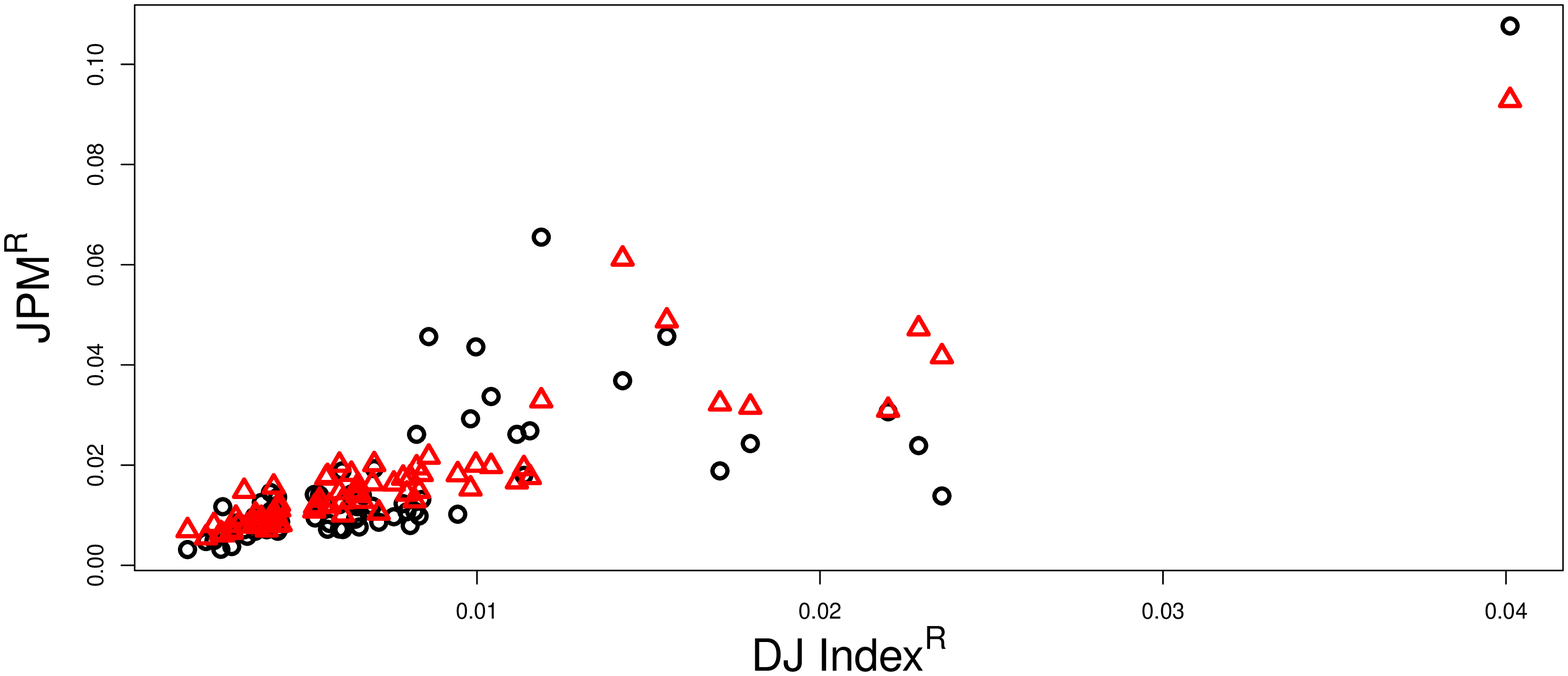} 
	\includegraphics[width=170pt, height=110pt]{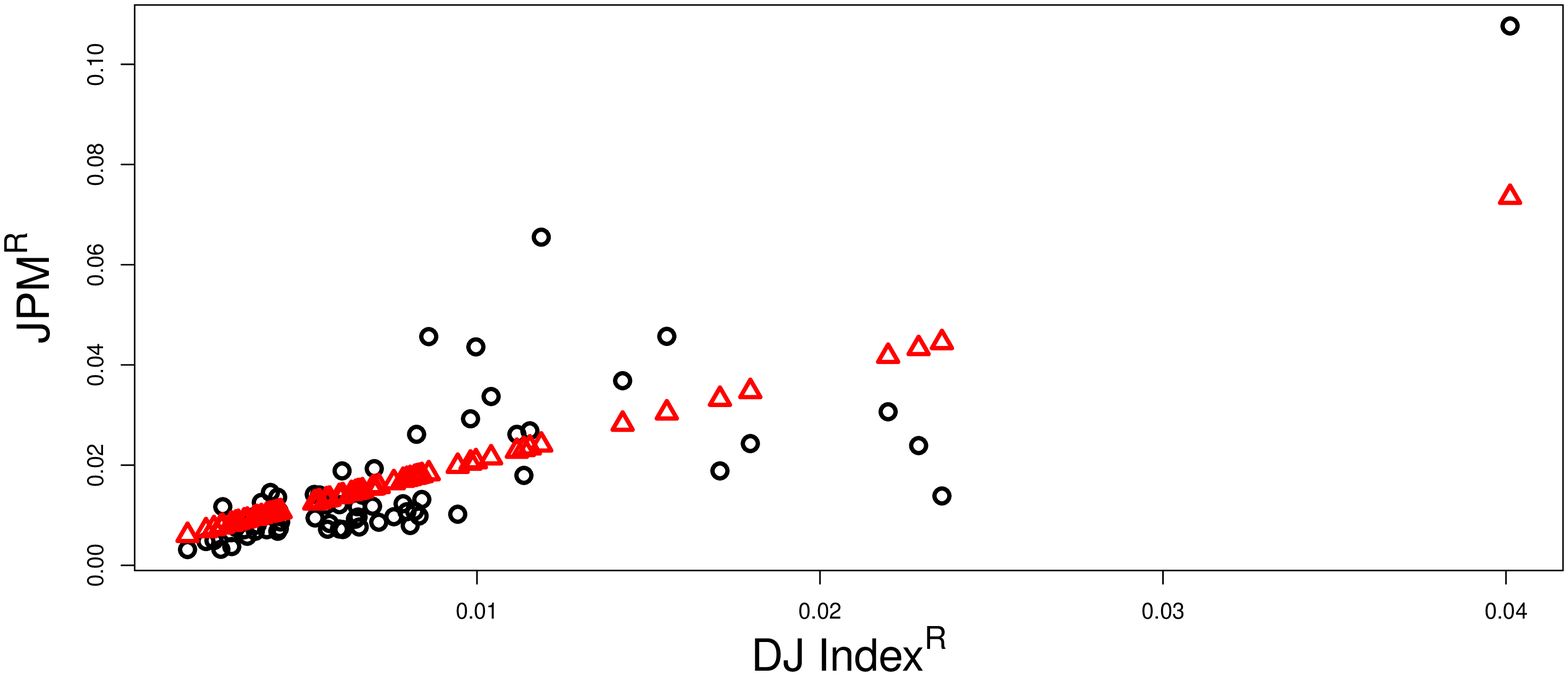}
	\caption{Overlaid plots of predicted centers and radii against the observed values for JPM. Observed data from the test set are represented by black circles and the predicted data are represented by red triangles.}
	\label{JPM_predicted}
\end{figure}

Predictive results for JPM price ranges are plotted in figure \ref{JPM_predicted} for illustration. For the predicted centers of BA, GE and JPM, random forests achieve a consistently better performance compared to CCRM. As for the radii, CCRM obtains slightly better accuracy by 0.04 for the $R^2$ of BA due to its relatively strong linear relationship with the DJIA index. For GE and JPM, random forests predict the data more accurately, compared to CCRM. Based on these results, the DJIA index is capable of predicting the three stock price ranges, accounting for a variance from 63\% to 97\% for the center, and from 64\% to 82\% for the radius, by random forest regression. The comparisons to CCRM show that random forests generally outperform CCRM in terms of predictive accuracy.

The three selected stocks, namely BA, GE and JPM, are leading companies in aviation, manufacturing and financial industries, so these stocks together may explain a good portion of variability of the DJIA index. In the second part of our analysis, we build predictive models for the DJIA based on the three stocks jointly. The multivariate analysis is run using both random forests and CCRM. As in the previous part, random forests include both the centers and radii of the stocks as the predictors for both the center and radius of the DJIA index. As the results show in table \ref{tab:center_table} and table \ref{tab:range_table}, for this multivariable analysis, random forests achieve better performance than CCRM. This is partly due to the extra predictor variables included in the random forests regression equations. In addition, the ability of random forests to account for nonlinearity is an advantage over CCRM.  

\begin{table}[htbp]
	\centering
	\caption{Predictive Accuracy for Asset Prices: Center}
	\begin{tabular}{lrrrrrr}
		\toprule
		& \multicolumn{2}{c}{$R^2$} & \multicolumn{2}{c}{MSE} & \multicolumn{2}{c}{MAE} \\
		\cmidrule{2-7}          & \multicolumn{1}{l}{CCRM} & \multicolumn{1}{l}{RF} & \multicolumn{1}{l}{CCRM} & \multicolumn{1}{l}{RF} & \multicolumn{1}{l}{CCRM} & \multicolumn{1}{l}{RF} \\
		\midrule
		BA    & 0.8655 & \textbf{0.9005} & 0.0089 & \textbf{0.0066} & 0.0806 & \textbf{0.0655} \\
		&       &       &       &       &       &  \\
		GE    & 0.5946 & \textbf{0.6259} & 0.0699 & \textbf{0.0644} & 0.2353 & \textbf{0.1880} \\
		&       &       &       &       &       &  \\
		JPM   & 0.5739 & \textbf{0.6904} & 0.0109 & \textbf{0.0079} & 0.0818 & \textbf{0.0655} \\
		&       &       &       &       &       &  \\
		DJIA Index & 0.8682 & \textbf{0.9665} & 0.0026 & \textbf{0.0007} & 0.0441 & \textbf{0.0185} \\
		\bottomrule
	\end{tabular}%
	\label{tab:center_table}%
\end{table}%

\begin{table}[htbp]
	\centering
	\caption{Predictive Accuracy for Asset Prices: Radius}
	\begin{tabular}{lrrrrrr}
		\toprule
		& \multicolumn{2}{c}{$R^2$} & \multicolumn{2}{c}{MSE} & \multicolumn{2}{c}{MAE} \\
		\cmidrule{2-7}          & \multicolumn{1}{l}{CCRM} & \multicolumn{1}{l}{RF} & \multicolumn{1}{l}{CCRM} & \multicolumn{1}{l}{RF} & \multicolumn{1}{l}{CCRM} & \multicolumn{1}{l}{RF} \\
		\midrule
		BA    & \textbf{0.6785} & 0.6357 & \textbf{1.87E-05} & 2.11E-05 & \textbf{0.0031} & 0.0032 \\
		&       &       &       &       &       &  \\
		GE    & 0.5311 & \textbf{0.7196} & 7.44E-05 & \textbf{4.45E-05} & 0.0043 & \textbf{0.0038} \\
		&       &       &       &       &       &  \\
		JPM   & 0.6522 & \textbf{0.7150} & 7.91E-05 & \textbf{6.48E-05} & 0.0059 & \textbf{0.0054} \\
		&       &       &       &       &       &  \\
		DJIA Index & 0.7708 & \textbf{0.8164} & 1.08E-05 & \textbf{8.69E-06} & 0.0021 & \textbf{0.0020} \\
		\bottomrule
	\end{tabular}%
	\label{tab:range_table}%
\end{table}%

\section{Conclusion}\label{sec:conclude}
We proposed random forest regression for interval-valued data. Its nonparametric nature automatically ensures mathematical coherence without any constraints, making it more flexible than most of the existing regression methods such as CCRM.  In addition to linear problems, it also handles nonlinearity very well. As most nonparametric methods require extensive tuning to achieve optimal performance, random forests are especially user-friendly in the sense that very little tuning is needed. Furthermore, they are robust against overfitting and high-dimensionality. All these advantages of random forest regression make it a potentially very promising method to be extended to interval-valued data. In this paper, we adopted bivariate regression. Simulation results confirmed its advantages over CCRM and the kernel estimator. Future research includes regression with multivariate response, namely, center and radius jointly, to account for their potential correlation.



\pagebreak
\appendix
\section{Random Sets Preliminaries}\label{append:preliminary}
Denote by $\mathcal{K}\left(\mathbb{R}^d\right)$ or $\mathcal{K}$ the collection of all non-empty compact subsets of $\mathbb{R}^d$. The Hausdorff metric 
\begin{equation*}
  \rho_H\left(A,B\right)=\max\left(\sup\limits_{a\in A}\rho\left(a,B\right), \sup\limits_{b\in B}\rho\left(b,A\right)\right),\ \forall A,B\in\mathcal{K},
\end{equation*}
where $\rho$ denotes the Euclidean metric, defines a metric in $\mathcal{K}\left(\mathbb{R}^d\right)$. As a metric space, $\mathcal{K}\left(\mathbb{R}^d\right)$ is complete and separable (\cite{Debreu67}).  In the space $\mathcal{K}$, a linear structure can be defined by Minkowski addition and scalar multiplication as
\begin{equation}\label{def:int-linear}
  A+B=\left\{a+b: a\in A, b\in B\right\},\ \ \ \ \lambda A=\left\{\lambda a: a\in A\right\},
\end{equation}
$\forall A, B\in\mathcal{K}$ and $\lambda\in\mathbb{R}$. However, $\mathcal{K}\left(\mathbb{R}^d\right)$ is not a linear space (or vector space), as there is no inverse element of addition.

Let $(\Omega,\mathcal{L},P)$ be a probability space. A random compact set is a Borel measurable function $A: \Omega\rightarrow\mathcal{K}$, $\mathcal{K}$ being equipped with the Borel $\sigma$-algebra induced by the Hausdorff metric. If $A(\omega)$ is convex almost surely, then $A$ is called a random compact convex set. (\cite{Molchanov05}, p.21, p.102.) The collection of all compact convex subsets of $\mathbb{R}^d$ is denoted by $\mathcal{K}_{\mathcal{C}}\left(\mathbb{R}^d\right)$ or $\mathcal{K}_{\mathcal{C}}$. Much of the random sets theory has focused on compact convex sets (\cite{Artstein75}, \cite{Aumann65}, and \cite{Lyashenko82}, \cite{Lyashenko83}). Especially, when $d=1$, $\mathcal{K}_{\mathcal{C}}(\mathbb{R})$ contains all the non-empty bounded closed intervals in $\mathbb{R}$. A measurable function $[X]: \Omega\rightarrow\mathcal{K}_\mathcal{C}\left(\mathbb{R}\right)$ is called a random interval. The expectation of a random compact convex random set $A$ is defined by the Aumann integral of set-valued function (see Aumann (1965), Artstein and Vitale (1975)) as
\begin{eqnarray*}
  E\left(A\right)=\left\{E\xi:\xi\in A \text{ almost surely}\right\}.
\end{eqnarray*}
In particular, the Aumann expectation of a random interval $[X]$ is given by
\begin{equation}\label{def:aumann}
  E\left([X]\right)=[E\left(X^L\right),E\left(X^U\right)]. 
\end{equation}
For each $X\in\mathcal{K}\left(\mathbb{R}^d\right)$, the function defined on the unit sphere $S^{d-1}$:
\begin{equation*}
  s_X\left(u\right)=\sup_{x\in X}\left<u, x\right>,\ \ \forall u\in S^{d-1}
\end{equation*}
is called the support function of X. Let $\mathcal{S}$ be the space of support functions of all non-empty compact convex subsets in $\mathcal{K}_{\mathcal{C}}$. Then, $\mathcal{S}$ is a Banach space equipped with the $L_2$ metric 
\begin{equation*}
  \|s_X(u)\|_2=\left[\int_{S^{d-1}}|s_X(u)|^2\mu\left(\mathrm{d}u\right)\right]^{\frac{1}{2}},
\end{equation*}
where $\mu$ is the normalized Lebesgue measure on $S^{d-1}$. According to various embedding theorems (\cite{Radstrom52}; \cite{Hormander54}), $\mathcal{K}_{\mathcal{C}}$ can be embedded isometrically into the Banach space $C(S)$ of continuous functions on $S^{d-1}$, and $\mathcal{S}$ is the image of $\mathcal{K}_\mathcal{C}$ into $C(S)$. Therefore, $\delta\left(X, Y\right):=\|s_X-s_Y\|_2$, $\forall X, Y\in\mathcal{K}_\mathcal{C}$, defines an $L_2$ metric on $\mathcal{K}_\mathcal{C}$. It is well known that $\rho_H$ and $\delta$ are equivalent metrics, but $\rho_H$ is less preferred for statistical analysis for several reasons. As an important one, $E\rho_H^2\left(X, h(X)\right)$ is not minimized at $h(X)=E(X)$.
Particularly the $\delta$-metric for an interval $[x]\in\mathcal{K}_{\mathcal{C}}(\mathbb{R})$ is
\begin{equation*}
  \|[x]\|_2=\|s_{[x]}(u)\|_2=\frac{1}{2}\left(x^L\right)^2+\frac{1}{2}\left(x^U\right)^2=\left(x^C\right)^2+\left(x^R\right)^2,
\end{equation*}
and the $\delta$-distance between two intervals $[x], [y]\in\mathcal{K}_{\mathcal{C}}(\mathbb{R})$ is 
\begin{eqnarray*}
  \delta\left([x], [y]\right)&=&\left[\frac{1}{2}\left({x}^L-{y}^L\right)^2+\frac{1}{2}\left({x}^U-{y}^U\right)^2\right]^{\frac{1}{2}}\\
  &=&\left[\left(x^C-y^C\right)^2+\left(x^R-y^R\right)^2\right]^{\frac{1}{2}}.
\end{eqnarray*}
\cite{Gil01} generalized the $\delta$-distance to the $W$-distance as
\begin{equation}\label{def:w-distance}
  d_W\left([x],[y]\right)=\left\{\int_{[0,1]}\left[f_{[x]}(\lambda)-f_{[y]}(\lambda)\right]^2dW(\lambda)\right\}^{\frac{1}{2}}, 
\end{equation}
where $f_{[x]}(\lambda)=\lambda x^U+(1-\lambda)x^L$, $\forall \lambda\in [0, 1]$, and $W$ is any non-degenerate symmetric measure on $[0, 1]$. The advantage of the $W$-distance lies in its flexibility to assign weights to the points in the interval. In particular, this can be interpreted as a probability distribution for a random point inside the interval. On the other hand, it can be shown that
\begin{equation}\label{def:w2}
   d_W^2\left([x],[y]\right)
   =\left(x^C-y^C\right)^2+\left(x^R-y^R\right)^2\int_{[0,1]}\left(2\lambda-1\right)^2dW(\lambda).
\end{equation} 
Notice that $\int_{[0,1]}\left(2\lambda-1\right)^2dW(\lambda)\in[0, 1]$ is a constant determined by $W$. So the $W$-distance can also be interpreted as choosing a weight for $\left(X^R-Y^R\right)^2$ in calculating the $L_2$ distance. 



\bibliographystyle{apa}
\bibliography{intervalregression}

\end{document}